# Billiards correlation functions[?]

Pedro L. Garrido[#], Giovanni Gallavotti[*]

*dedicated to Philippe Choquard on his sixtyfifth birthday*

*Abstract*: We discuss various experiments on the time decay of velocity autocorrelation functions in billiards. We perform new experiments and find results which are compatible with an exponential mixing hypothesis, first put forward by Friedman and Martin, [FM]: they do not seem compatible with the stretched exponentials believed, in spite of [FM] and more recently of Chernov, [C], to describe the mixing. The analysis led us to several byproducts: we obtain information about the normal diffusive nature of the motion and we consider the probability distribution of the number of collisions in time $t_m$ (as $t_m \to \infty$) finding a strong dependence on some geometric characteristics of the locus of the billiards obstacles.

Keywords: Billiards, correlation functions, velocity autocorrelation, diffusion coefficients, Lorentz model, mixing, ergodic theory, chaos, Lyapunov exponents, numerical experiments.

[?] This paper is archived in the electronic archive mp_arc: send an empty e-mail message to `mp_arc@math.utexas.edu` to get the instructions on how to copy a postscript version of this preprint.

[#] Departamento de Fisica Moderna, Universidad de Granada, 18071 Granada, Spain

[*] Dipartimento di Fisica, Università di Roma, P.le Moro 2, 00185 Roma, Italy

## §1 *Billiards correlation functions.*

The first results on the ergodicity of the billiards go back about thirty years ago, [S]. But, as it is well known, ergodicity is a very weak, and in some sense, not too interesting, property. More direct physical meaning is attached to the correlation functions and to their decay speed.

Let $\Omega$ be the system phase space, $S_t$ the evolution map, $\mu(dx)$ the Liouville measure.

In the case of the billiards $\Omega$ is three dimensional: a point $x \in \Omega$ is a point $q \in M$, where $M$ is the billiards *table*, see Fig. 1, §3, *i.e.* a periodic box of side 1 with a few circles of radii $R_1, R_2, \ldots$ taken out and regarded as *obstacles*, and an angle $\varphi \in [0, 2\pi]$ so that $(q, \varphi)$ represents the position and the velocity direction (with respect to the $x$-axis, say) of a point mass moving with unit velocity and direction $\varphi$ until a collision takes place with the obstacles. Upon collision $\varphi$ changes according to the elastic collision rules (equal incidence and reflection angles). The measure $\mu(dx)$ is simply $\mu(dx) = d^2q\, d\varphi/(2\pi|M|)$, where $|M|$ = area of $M$. The dynamical system $(\Omega, S_t, \mu)$ will be called the *continous* or $3d$ *system* (because it has a 3 dimensional phase space). The average with respect to $\mu$ over the phase space $\Omega$ will be denoted by the symbol $\langle . \rangle$.

An associated dynamical system is the *collision system* or $2d$ *system*: its (2 dimensional) phase space consists of points $q \in \partial M$ and of the *incidence* angles $\vartheta \in (\frac{\pi}{2}, \frac{3\pi}{2})$ formed by the velocity at collision and the outer normal to the obstacle, counted counterclockwise. We denote $(\Omega_c, S^c, \mu_c)$ the "collision dynamical system" with $\Omega_c$ being the phase space, $S^c$ being the map mapping one collision $\xi = (s, \vartheta, i)$ to the previous collision $\xi' = (s', \vartheta', i')$, where $i$ is a label for the obstacle on which the collision takes place, $s$ is the arc of the obstacle point $q$ where the collision takes place (counted from an arbitrarily fixed origin), and $\vartheta_c$= collision angle, and $\mu_c$ is the invariant measure: $\mu_c(d\xi) = -\cos\vartheta\, d\vartheta\, ds/$normalization. The average with respect to $\mu_c$ over the phase space $\Omega_c$ will be denoted by $\langle . \rangle_c$ (and no confusion should arise).

We shall consider three types of billiads (*i.e.* three configurations of obstacles): $\infty$H, 0H and D, defined precisely below.

By *correlations decay* one does not mean the behaviour as $t \to \infty$ of:

$$\langle f(t) f(0) \rangle = \int \mu(dx) f(S_t x) f(x) \tag{1.1}$$

for the most general *observable* $f$, *i.e.* for the most general measurable function $f$ on phase space. One is, in fact, interested in the (1.1) only for *very special* observables $f$. On the other hand it is clear that, the systems $(\Omega, S_t, \mu)$ and $(\Omega_c, S^c, \mu_c)$ being isomorphic to Bernoulli schemes, [GO], one can find nasty functions $f$ for which (1.1) approaches $\langle f \rangle^2$ as slowly as wished (by those who care about such wishes).

A class (almost exhaustive) of *interesting* observables and related quantities is in the following list.

1) the *$x$ component of the velocity* $v_x = \cos\varphi$ in the dynamical system $(\Omega, S_t, \mu)$, leading to the *velocity autocorrelation* function:

$$C(t) = \langle v_x(t) v_x(0) \rangle \tag{1.2}$$

where $v_x(t)$ is the velocity at time $t$ of an initial datum with velocity $v_x(0)$.

2) the *$x$ component of the collision velocity*, in the collision dynamical system $(\Omega_c, S^c, \mu_c)$, $v_{cx} = \cos\varphi$ where $\varphi = \vartheta - \varphi_q$, $\varphi_q$ being the angle between the normal at the collision point $q$ and the $x$-axis, $(-\pi/2 \le \varphi_q \le \pi/2)$. This observable leads to the *collision velocity autocorrelation* function:

$$C_c(n) = \langle v_{c\,x}(n) v_{c\,x}(0) \rangle_c \tag{1.3}$$

where $v_{cx}(n)$ is the velocity at the $n$-th collision of an initial datum with velocity $v_{cx}(0)$.

3) the *transversal velocity autocorrelation* given, with obvious notations, by:

$$C^T(t) = \langle v_x(t) v_y(0) \rangle \tag{1.4}$$

Other related quantities are:

4) the *square displacements* $s(t)$ and $s_c(n)$, associated with the billiards and the collisions systems, respectively,

defined by:

$$\begin{aligned} s(t) =& \langle x^2(t)\rangle \equiv 4\int_0^t d\tau \int_0^\tau d\tau' \langle v_x(\tau)v_x(\tau')\rangle \\ s_c(n) =& \langle x^2(n)\rangle_c, \qquad D = \lim_{t\to\infty} \frac{s(t)}{t} \end{aligned} \tag{1.5}$$

where $x(t), x(n)$ are two dimensional vectors which are measured by imagining to unfold on the plane the periodic box into all its images. And by thinking the motion to take place, without periodic boundary conditions but in the full plane, among the periodic lattice of scatterers generated by the obstacle images. In this case it is clear that $D$ can be called the *diffusion coefficient* and the mean square displacement is, with the above notations, $Dt$ (hence, for the purpose of comparison with [Bl] our diffusion coefficient $D$ is related to the one, which we call $D_0$, in [Bl] eq. (1.9) by $D = 2\sqrt{2}D_0$).

5) the number of collisions, $\nu_{t_m}(x)$, undergone by the motion starting at $x$ in the time $t_m$. This is a random quantity if $x$ is chosen randomly with distribution $\mu$ (or $\mu_c$); then we can ask the value of the probability distribution:

$$d_\alpha(z) = \lim_{t_m\to\infty} \frac{1}{\delta z} \text{prob}\left((\nu_{t_m} - \langle\nu_{t_m}\rangle)t_m^{-\alpha} \in (z, z+\delta z)\right) \tag{1.6}$$

where $\langle\nu_{t_m}\rangle$ is the average of $\nu_{t_m}(x)$ over $x$ and $\alpha > 0$.

The theoretical analysis of the above quantities started becoming at least conceivable after the work of [BS], about twelve years ago.

The possibility of constructing a *symbolic dynamics* representation of the billiards motion, [BS], permitted to obtain the first upper bounds of the decay of $C_c(n)$.

The bounds have been studied in the 0H case (defined precisely below) and have the form:

$$|C(t)| \le ae^{-b(t/t_0)^\varepsilon}, \qquad \text{or} \qquad |C_c(n)| \le e^{-bn^\varepsilon} \tag{1.7}$$

for some $\varepsilon > 0$ (small), $a, b > 0$, where only the second bound is really proved in [BS].

The above (1.7) is just an upper bound and, very soon after [BS], the matter started being investigated numerically. The first results seemed to suggest that $C(t)$ really did behave, on the sequence of times where the local maxima of $|C(t)|$ are achieved, as a streched exponential with $\varepsilon$ ranging between 0.4 and 0.8 depending on the scatterers geometry, [CCG], [BlD]. A *stretched exponential* is a function of $t$ or $n$ of the form (1.7) and $\varepsilon$ is called the *stretching. With the notable exception of [FM] who clearly stated that, in the 0H case (see below), their results indicated an exponential decay of the sequence of local maxima*, consistent with a representation of $C(t)$ as a product of a pure exponential times a periodic function.

Here by *geometry* of the scatterers one means some rough qualitative property: namely whether there are collisionless trajectories (billiards *with horizon*, denoted $\infty$H), or not (billiards *without horizon*, denoted 0H), when the periodic array consisting of the images of the obstacles does not allow to draw a path to infinity avoiding the obstacles, or *diamond* billiards, denoted $D$, in the other case where the obstacles keep the particle inside a bounded region.

On the other hand the reason why one expects a stretched exponential decay seems to be related to the fact that the Markov partition realizing (see [BS]) the symbolic dynamical representation is not finite but countable.

The denumerability of the Markov partition depends on the basic lack of smoothness of the billiards $(\Omega_c, S^c, \mu_c)$ system dynamics: there are sharp discontinuities on the one dimensional lines corresponding to collisions which are preceded by a tangent collision (recall that we look at the motion backwards in time).

If the Markov partition had been finite and the dynamical system smooth, the hyperbolicity of the billiards would have allowed us to conclude that the autocorrelation of any smooth observable $f$ would have approached its $n \to \infty$ (infinite time) limit exponentially fast, in particular $C_c(n)$ would have been exponentially decaying (and hence presumably also $C(t)$ because, under such (hypothetical) assumption, the time between collisions should be expected to be finite with finite moments). This is in fact a general mixing property for smooth observables in smooth hyperbolic systems, see [S2,Bow].

From the theory of the Markov partitions in [BS] one expects the partition to be "essentially finite", *i.e.* to consist of sets of apparently regular shape (irregularities breaking in two the elements of the partition, like tiny cuts, should be confined to very small scales), a finite number of which fills, for all practical purposes, the phase space. Furthermore, and most important, one expects to be able to construct "most" of the elements of the partition by examining only the first few successive collisions of general initial data, at least in the cases of 0H and D billiards, for which the time between collisions is finite with all its moments.

If this is correct, as preliminary evidence deduced from our attempts to device a workable numerical algorithm to construct the Markov partition (in the 0H case) seems to suggest, one would expect that the smooth observables like the velocity $f = v_x$ mix exponentially fast in the 0H and D billiards: this means that in order to see non exponential behaviour (like stretched exponentials or even long time tails) one would need extremely accurate experiments over very many collisions. Otherwise the system should behave as a normal hyperbolic system without singularities. This is essentially our point of view (it could be called a "conjecture" but we refrain from formalizing it).

In the case of $\infty$H billiards, the same remarks should apply to the collision correlations, *i.e.* the correlations in the $2d$ collision dynamical system, even though the expected time between collisions has divergent moments of any order $\geq 2$. The correlations in the $3d$ dynamical system will have slow time decay (see [Bl]) as a consequence of the fact that there is also an infinite expectation value for the time of the first collision (as the initial velocities generating motions not experiencing collisions before a time $t$ are those in contained an angle of order $t^{-1}$). Note however that such a long time tail might become visible only for $t$ so large to be not observable (with the present computer tools): and in fact this appears to be the case (see below).

The above picture seems in sharp contrast with most of the existing numerical experiments, with the exception of [FM]. It does however agree with the theoretical work of [C] where a similar, but far simpler, non smooth hyperbolic system has been considered and exponential time decay of the correlations has been shown, in spite of the non existence of finite Markov partitions. The results of [C] led in fact Chernov to conjecture (independently) the exponential decay for the billiards collision systems as well.

Having developed large sets of data, and seeing still quite far ahead the development of a numerical construction of a Markov partition, we thought that it would be useful to use our data to check the consequences of the above viewpoint on the correlation decay. Therefore we undertook a few extra experiments to see how strong was the evidence for stretched exponentials, and if one could confirm the important results of [FM], which had not received the deserved attention: unfortunately we became aware of this work too late for using exactly their triangular lattice configurations and to perform a detailed quantitative comparison. Our attempts at constructing Markov partitions had already led us to special billiard configurations which are not exactly the one of [FM]. Our work (being made about eight years after [FM]) contains better statistics (as computers are larger), but it is compatible with theirs.

It should, however, be made clear, at the start, that *in absence of rigorous asymptotics* it is only possible to check if the experimental results are compatible with some *a priori* given function describing the decay. Thus it makes sense to test whether the experimental results can be fitted with a function of the form: $e^{-at}f(wt)$ with $f$ periodic. This implies that the successive maxima are placed on one or more parallel straight lines on a logarithmic plot. Of course one could test more complicated functional dependences (e.g. quasi periodic, aymptotically quasi periodic or stretched exponential times a periodic function, etc.), but the best one can expect from (our) analysis of the experimental results is a check of compatibility with an assumed law. If the assumed law allows for many parameters it is likely to give apparently better results; for instance a fit with: $e^{-at^\alpha}f(wt)$ and $f$ periodic, could give slightly better fit than the one with $\alpha = 1$: this can be interpreted if, for instance, the best fit over $(a, \alpha)$ yields $\alpha$ close to 1, to mean that $\alpha$ is actually 1 and $f(t)$ is close to periodic, with some small nonperiodic corrections.

This difficulty with the analysis of experimental data is particularly evident in the case of the collision correlation function, $C_c(n)$, where the observations can only be made on integer values of the argument and even a periodic law would not necessarily yield data lying on parallel lines on a logarithmic plot.

The following description of our experimental results seems to confirm the findings of [FM] and provides, in our opinion, further evidence for a purely exponential time decay of the correlation functions $C(t), C_c(n)$ as well as a diffusive behaviour of the motion (unfolded among the lattice of the images of the scatterers) when the scatteres do not confine the motion to a finite region (billiards $\infty$H or 0H, *i.e.* non diamond billiards). There are some puzzling features that we could not resolve, see comments to fig. 6. As a byproduct we studied also:

1) The distribution of the number of collisions in a given time $t_m$, for $t_m$ large, finding qualitative differences

between the $\infty$H billiards and the 0H or diamond billiards.

2) The diffusion phenomena in the $\infty$H and the 0H cases.

We shall (of course) argue that the "unexpected" exponential decay of the $C(t)$ correlation function in the $\infty$H case, and the associated diffusive behaviours is an artifact, because the long time tail is not yet visible on our computer experiment time scale.

The diffusion seems "normal" (with mean square displacement of order $t$ as $t \to \infty$) in the 0H case while is very likely (see [Bl]) anomalous (or "superdiffusive" with mean square displacement of order $t \ln t$) in the $\infty$H case: but such asymptotic regime does not seem to be visible at the time scales that we reach.

Furthermore in the first case the distribution of the number of collisions occurring in a given time appears to be an asymptotically gaussian distribution. In the second case the distribution of the number of collisions appears to decay exponentially fast to the right and very sharply to the left of some value $z_0$, although we again think that the latter is probably a short time effect.

Our work seems to provide evidence for exponential decay of $C(t)$ and $C_c(n)$: but the precision does not rule out exponential stretching with $\varepsilon$ close to one. *A fortiori* logarithmic stretching like:

$$|C(t)| \le a e^{-b(t/t_0)(\ln t/t_0)^{-\varepsilon}} \tag{1.8}$$

is not ruled out either, and in fact it appears to us to be a natural candidate. But this can probably come out only from an analytic theory.

The above discussion expresses numerical results obtained by experimenting over sequences of many collisions (more than the previous experiments) but still not in very large numbers (the hyperbolic nature of the motion prevents, even by performing double precision computations, to study more than about 20 collisions in the D case, 15 in the $\infty$H case and 10 collisions in the 0H case, see fig.5, for instance). What really happens beyond such time scales is not at present analyzable numerically, at least not in the sense of the present paper (and not much improvement can be expected in the future, even in the far future, as we are dealing with a chaotic system). Analytical work is here more promising, but of course difficult (see in particular [Bl]).

One final comment: in order to avoid just repeating the work [FM] with better statistics we tried organize our data in such a way that they could be reliably reproduced (by those interested in them), including the error bars and the statistical analysis. This is explained in more detail in the next section, and it is the reason why we cannot consider very long time intervals. *Remarks on numerical experiments.* At the editors request we add a few general thoughts on numerical experiments on asymptotic properties of dynamical systems. There seems to be still concern on the actual measurability of quantities like Lyapunov exponents, decay of correlations, transport coefficients in numerical experiments. The problem seems to be that such quantities are defined by limits and "computers' cannot take limits". In fact the same argument could be construed for actual laboratory experiments: the latter seem to bother people less, as a true laboratory is far more reassuring than a computer, because it is older and we are used to it. It is important stressing that "computer experiments" are just experiments, worth of attention on their own; and the interpretation of their results may rest on theoretical frameworks which are more idealized. Measuring a Lyapunov exponent on a computer has the same meaning as measuring the period of the orbit of Jupiter, or the inclination of Mars. We measure them very accurately, but we are not sure that the motion is actually periodic, nor that the inclination is constant (they are not! but they "appear to be"). But in theoretical mechanics there are models of planetary motions in which Jupiter is moving periodically, or "can", and the Martian seasons are forever as dull as they are now. This gives us the "concept" of period and of inclination, and the idea of mesuring them, and of testing if they are constant. But of course our check of the constancy cannot be done as our lifes are too short (and the age of the universe is also too short): nevertheless we perform the measurements as accurately as possible and come out with precise figures. The figures tell us that, performing the measurements we did to measure the period or the inclination "as if they were well defined and constant", we get some definite results (by the way this may not be so easy).

The interpretation of the results is customarily given in such a form: the period of Jupiter is $T = \ldots \pm ..\%$ and the average inclination of Mars is $i = \ldots \pm ..\%$. Of course it is understood that they may change, or that they may simply not exist. But such doubt is not stated, usually; certainly not in the case or Mars' inclination. Hence this means that, on the basis of some theoretical models such motions are possible, and the data that we observe *are consistent* with the above values of $T, i$. Since the values are reproducible they qualify as "good data". What else could we possibly want? In the same sense in dynamical systems there are concepts like mixing, ergodicity, Lyapunov exponents which are all asymptotic: they are well defined in abstract, idealized, models . We can however perform the operations that it would be necessary to measure them, get some results and then call such results with the corresponding names. We cannot check, in measuring Jupiter's period, that it returned after the jovian year exactly in the same position and not $1.mm$ away (which could imply a double value of the period, and an infinite time to really check); likewise we do not check that the values of the Lyapunov exponents do not change by really going to a limit of infinite time or to twice the maximum time that we can reach (controlling the errors that we (think we) make). The numbers that we get are the "effective" Lyapunov exponents. In fact they are a far more interesting quantity than the abstract, non measurable, Lyapunov exponents: they are the ones "we would feel" if we had to interact with the system on the experimental time scale. The same can be said about the correlation functions. Of course the natural evolution of human affairs will perhaps make longer time scales experimentally reachable and new experiments to be performed. In fact it is now known that the average inclination of the axis of Mars is not fixed at all, it greatly moves, randomly (by about as much as $50^o$!), with a Lyapunov exponent of about the inverse of 1 million years, [LR]: which means that its motion might well be periodic (and with zero Lyapunov exponent) but for all practical purposes it is not such on rather short time scales (the result emerges from computer experiments, but once we finally succeed in being permitted to land on Mars, by the apparently unwilling inhabitants, we shall probably find geological evidence for that).

In this paper we "measure correlations" in billiards. The deep theorems that are available do not give many hints, if any, on how to measure reliably (*i.e.* with an *a priori* controlled error) such quantities; in some cases there are not even theorems concerning their very existence, not to speak of approximability with finite algorithms (which is the goal, or should it be?, of reasonable theoreticians wishing to do scientific research and not philosophical speculations). Hence the real definition of the quantities is the one which emerges from the experiment that we describe. The minimum requirement is that it should be reproducible, and as accurate as possible: the language in which the results are formulated is necessarily borrowed from some general framework or "formalism", (ergodic theory and billiards theory in the present case). But they are not, and cannot, be theorems. They are just facts that hopefully, if the experiment is meaningful, may give new ideas to the theory and form a more satisfactory picture of the phenomena. But it is illusory that we can ever prove any theory by an experiment and viceversa: it is in fact incomprehensible to us why nevertheless a close connections between reality and theory can be at all empirically established. It might be because of what Galileo noted as the book of nature being written in mathematical characters: but books may not tell the truth, which in fact may just be undefined and undefinable. What really remains after performing an experiment is the expectation that others will find it interesting and useful (and this can only be if it is reproducible; and in spite of reproducibility it will often not be regarded as interesting or meaningful).

We therefore use freely in what follows notions like correlations, distributions, decay, *etc.*: they are defined in the text, and ultimately in the present case by the computer program. And we tried to define clearly what we do, how we do estimate the errors; and we have avoided letting the computer just compute, trying to avoid the reader the frustation that we experienced many times in reading of numerical works (unquoted here) in which there was not enough information for reproducibility. We are, of course, aware that, for instance, the quantities like "diffusion coefficients", or "Lyapunov exponents", cannot in principle be measured from our data: but the values we provide for them are, for all practical purposes, the real ones in our experiment; and, if reproducible, in all experiments of the same kind.

We know of no numerical experiments on billiards or other dynamical systems (even smooth) in which, for instance, the Lyapunov exponents are numerically estimated over a time longer than allowed to expand the round off error beyond the precision of a few percent (*i.e.* the usually reported errors on such exponents ): up to 20 collisions in our case. Therefore we have limited the duration of our experiments to a few collisions: more data could have been collected, but we could not have reliably intepreted them. For our billiards there are just not big enough computers for many more collisions (and there will never be, as their required size grows "exponentially" with the number of collisions). This is the reason it would be better to perform our measurements by other methods: but we know of no other methods for the billiards case. They might be developed in the future. We think that the pioneering age is over and the computer experiments should be subject to no less stringent standards than "ordinary" experiments (because we think they are no different). Therefore error bars should be mandatory, the error analysis should not be left out, round off should not be ignored, *etc.*, and random number generators should be subject to tough tests. Many researchers do this and we would be happy to be considered at least close to them.

*Acknowledgements:* Work mainly supported by Rutgers University and the authors are indebted to Prof. J.L. Lebowitz for making it possible through grant NSF-DMR92-18903. Partial support is acknowledged to the italian agencies CNR-GNFM, Ministero della Ricerca and to the computer center of Istituto Astronomico (VAX4000/200) of the Università di Roma (G.G.), and mainly to the spanish agencies DGICYT (PB91-0709) and Junta de Andalucia (PAI) (P.G.). We are indebted to the referee for requesting that details about our random number generators be included.

§*2 The computer experiment.*

Our system is a square with periodic boundary conditions with sides of unit length, $a = 1$. We take the center of the torus as the origin of coordinates: $(0, 0)$. There are a circle of radius $R_1$ and center at $(0, 0)$ and four more circles with radius $R_2$ and centers at $(1/2, 1/2)$, $(1/2, -1/2)$, $(-1/2, -1/2)$, $(-1/2, 1/2)$. Obviously only the part of the circles inside the torus is relevant (see **Fig. 1**).

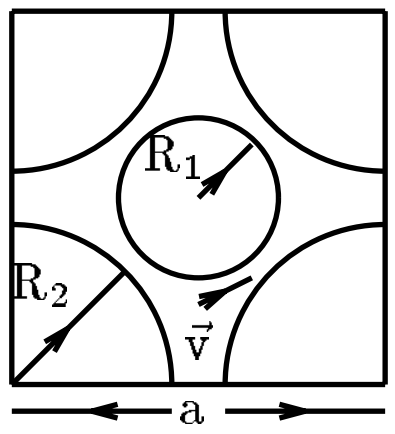

Fig. 1

*General billiard structure with scatterers of radius $R_1$ and $R_2$ in a box with side length a.*

A point particle is moving freely with unit velocity, $|v| = 1$, in the space external to the circles and hitting them elastically (conserving the modulus of the total momentum and the energy).

We studied three different cases (see Fig.1):

a) Billiards with infinite horizon ($\infty$H): $R_1 = R_2 = 7/20$, less than the maximal value $4^{-1}\sqrt{2}$, but larger than the value $1/4$ considered in [Bl], p.366 (where the length unit is $1/\sqrt{2}$, as the square lattice of the obstacles is with side 1 while in our case it is $1/\sqrt{2}$; the velocity is however 1 in both cases so that, calling $x(t)$ the position at time $t$, it is related to the position $x_B(t)$ of [Bl] by $x(t) = \frac{1}{\sqrt{2}} x_B(t\sqrt{2})$).

b) Billiards without horizon ($0H$): $R_1 = 1/5, R_2 = 2/5$.

c) Diamond ($D$): $R_1 = R_2 = (5/32)^{1/2} > 4^{-1}\sqrt{2}$.

The value for the radius chosen for the Diamond case is the same of [CCG] and it will make it easier the comparison with their results.

We use the following algorithm. The initial particle position is chosen at random with uniform distribution from the external space left by the circles. We give to the particle an initial velocity vector with an angle with respect to the horizontal axis chosen from a uniform distribution between 0 and $2\pi$. We compute whether the future particle trajectory will hit a circle or a box side and, in every case, the particle is moved up to one of those possibilities. If the particle hits a box side, periodic boundary conditions are applied and if it hits a circle we change the velocity direction according to the equations for an elastic collision with a surface with a given curvature. We apply the algorithm until the time of the particle evolution is equal to 4 units of time (an average of 20 collisions, as we shall see) and then we start again the algorithm.

*We do not apply molecular dynamics: the succesive collisions are determined by solving, on the computer, the appropriate equations for the geometric intersections.*

Note that for the 0H and D billiards, the above experiment permits us to study the collision correlations without performing new experiments. In fact, by choosing initial $3d$ data with distribution $\mu$ automatically produces a distribution $\mu_c$ of the first collisions. And we performed a few experiments measuring the collision (2d) correlation function even though they are very *difficult to interpret* (as commented in §1).

For the $\infty$H case, in contrast, by choosing initial $3d$ data it is not possible to produce the distribution $\mu_c$ because some collisions require to occur a time larger than our experiment. Therefore, in the case of $\infty$H, we have also done a set of computer simulations starting with an initial datum which is a collision (*i.e.* the position is randomly choosen on an obstacle boundary with uniform distribution and its direction is a collision angle $\vartheta$ chosen randomly between $\pi/2$ and $3\pi/2$ with distribution $-\cos\vartheta\, d\vartheta$). Hence in the latter case we study both, the $2d$ *system* and the $3d$ *system*. This is necessary because, as said above, for the $\infty$H case the average distance between collisions in the $3d$ system is $\infty$, while in the $2d$ case it is finite (although it has a divergent second moment). Therefore, *a priori*, the latter property could affect the time behavior of the macroscopic averages. On the other hand, it is very useful in itself to study the $2d$ case in order to compare with the most recent theoretical works on this billiard type (p.e. see [Bl]).

In all the graphs referring to the $3d$ systems the time is measured in units of:

$$t_0 = \frac{\int \tau(s,\vartheta) ds dc}{\int ds dc} \tag{2.1}$$

where $dc = -\cos\vartheta\, d\vartheta$ and $\tau(s,\vartheta)$ is the time between the collision $(s,\vartheta)$ and the previous one; so that $t_0$ is the theoretical average time between collisions (note that, by the ergodicity theorem [S], this *is exactly equal* to the average time between collisions in the $2d$ collision system as $dsdc$ is the invariant measure in the collision system defined at the begining of §1). In particular, we get: $t_0 = 0.151...$, $0.316...$ and $0.1645...$ for $D$, 0H and $\infty$H cases.

The plotted functions were the average over $2\times 10^7$ trajectories for the Diamond case, $1\times 10^7$ trajectories for $\infty$H case and up to $2.8\times 10^8$ trajectories for the 0H case. This implies a statistical error proportional to $N^{-1/2} \simeq 2\cdot 10^{-4}$ for $D$, $3\cdot 10^{-4}$ for $\infty$H and $6\cdot 10^{-5}$ for 0H.

In order to provide reproducible data and errors we describe in the Appendix our precise definition of the type of fit we do on the data, the evaluation of the errors on the fitting parameters and the criteria used to measure quantitatively how good different fits are. We think that our data are careful enough to be reproducible, errors included.

We were worried about the rounding error propagation due to the system chaoticity. This error is proportional to $s(0)e^{\lambda t}$ where $s(0)$ is the initial rounding error ($10^{-16}$, in our case) and $\lambda$ is the system maximum Lyapunov exponent. We know from our own computer simulation estimates that $\lambda \simeq 1.03\, t_0^{-1}$ for $D$ , $\lambda \simeq 1.75\, t_0^{-1}$ for 0H and $\lambda \simeq 1.23\, t_0^{-1}$ for $\infty$H. The original results existing in the literature are, in the $D$ case, due to [Be] and our computations agree within the errors with them.

Then, we may conclude that the statistical error dominates over the rounding error when $t$, measured in absolute units of time, is less than 4.2, 5.2 and 3.9 for $D$, 0H and $\infty$H cases respectively. Therefore, in our computer simulation (*going up to a time (measured in absolute units)* $\leq 4$), the only relevant source of errors should be the statistical ones. Computer simulations with longer time interval will measure hydrodynamical long time properties, i.e. time correlations between "different" initial and final trajectories.

*The error bars reported in the graphs are the mean square dispersions of each reported value added to an estimate of the other error sources (round off propagated by chaoticity). They are always plotted although most of the times they cannot be seen because of the graph width.*

Since we are interested in asymptotic quantities we do not want to include in our fits the short time data as they certainly show transient phenomena. Therefore we decided, quite arbitrarily, to discard the data produced in the first two units of time measured in units of the appropiate Lyapunov exponents. Such time scale, *Lyapunov time scale*, is different from the time scale used in the graphs (namely the (2.1)): therefore we marked in each graph the *Lyapunov time scale* by an arrow pointing at its value in the $x$-axis.

In the $3d$ system (*i.e.* continuous time dynamical system) we measured the magnitudes at times $t_i = i * t_m/4000$ ($i = 1, 2, .., 4000$), where $t_m$ is the *measured time interval* in absolute units. The computation of such $3d$ magnitudes is the most expensive part in CPU computer time because for *each trajectory and variable* we have to perform, at least, 4000 operations more than the usual ones for the $2d$ system (approximately 1500 operations in a trajectory of 15 collisions). We represent in the correponding graphs 400 points that we got by averaging locally 10 data points. The averaging is *done only to reduce the number of data plotted* and it does not differ appreciably from the full plot (*i.e.* the first it is within the error bars of the second) which represents the data really used in our fits. The symbols regularly used in the plots are small black circles (or ellipses) and big empty ones for the $3d$ and $2d$ functions respectively. The error bars appear (when visible) as vertical bars.

*§3 Billiards velocity autocorrelations:* $C(t) = \langle v_x(0)v_x(t)\rangle$.

a: The $\infty$H ($3d$) case:

In **Fig.2** we show the $|C(t)|$ behavior. We see the characteristic oscillations with monotonically decreasing amplitude. We manage to observe up to 6 or 7 oscillations before the statistical errors obscure the data. We realized how the amplitudes decrease with an apparently regular law and the oscillation period seems to be constant $\tau = 2.55(\pm 0.08)\, t_0$.

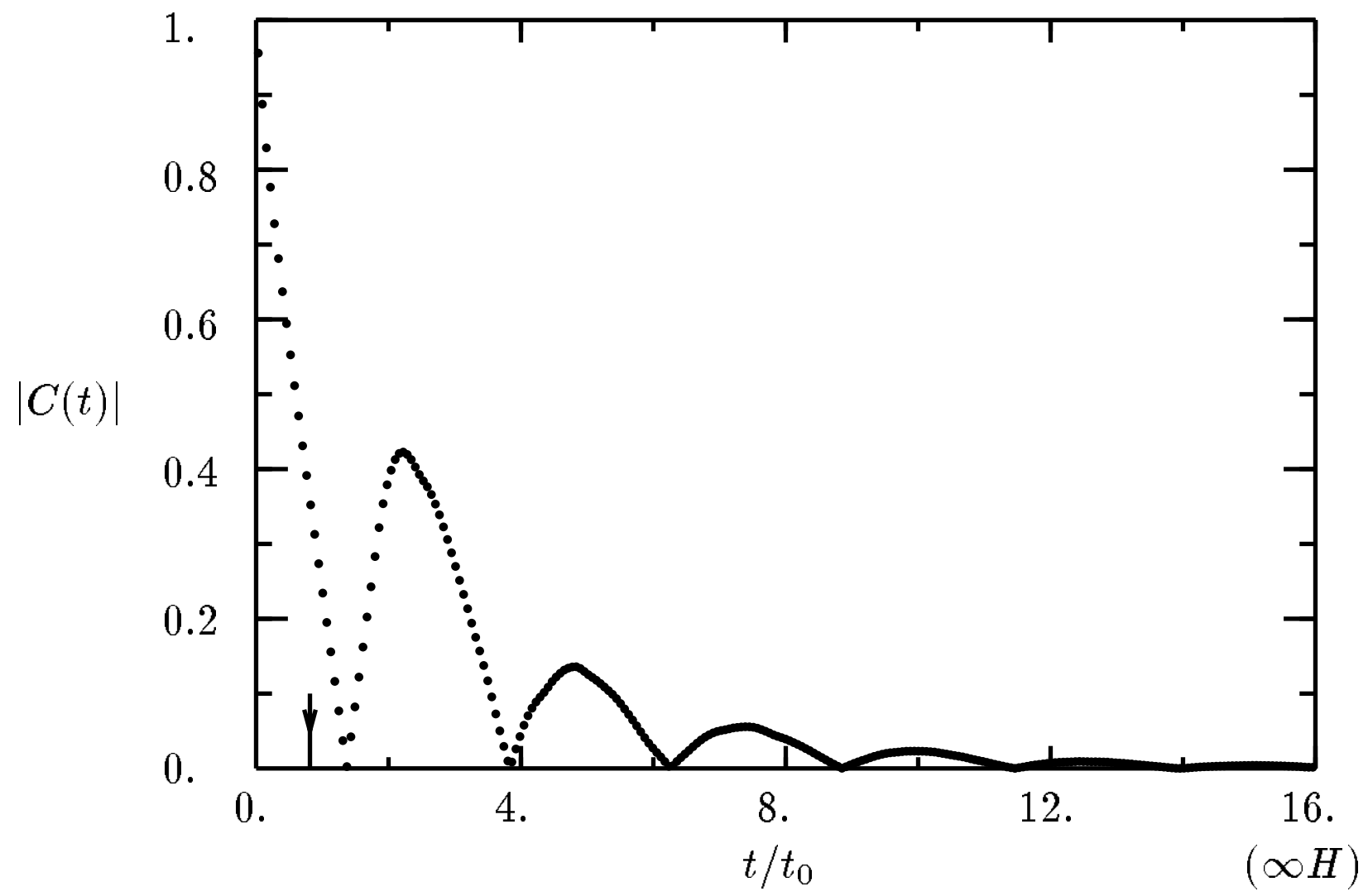

Fig. 2

*Absolute value of* $|C(t)|$*, the velocity-velocity correlation function, versus* $t/t_0$ *for the* $\infty H$ *case. The arrow marks the Lyapunov time scale (here and in the following pictures).*

Hence we try the fit of the data with a law $e^{-at/t_0} f(wt)$ with $f$ $2\pi$-periodic. If this is a good guess we should be able to get a good fit of the data $\ln|C(t)|$, evaluated on the relative maxima points $t_i$, with a function of the form: $-at/t_0 + b$, with $b = \ln f(wt_i)$. If the function $f$ is not too complicated, the time $t_i$ should appear at multiples of $2\pi w^{-1}$. More generally one could expect the maxima to occur on parallel lines of the form $-at/t_0 + b_i$, with $b_i = \ln f(\tilde{t}_i)$ and $\tilde{t}_i$ separated by $2\pi w^{-1}$.

This is confirmed when we plot the $\ln|C(t)|$ versus $t$ (see **Fig.3**). We discard the points with $t \le 3t_0 \approx 2\lambda^{-1}$ and we may fit a straight line crossing the last 5 *maxima*: $-0.34(\pm 0.05) - 0.345(\pm 0.006)t/t_0$. This implies a pure exponential law behavior for the amplitude with a decay rate $2.90(\pm 0.05)t_0$, apparently unrelated to the Lyapunov exponent.

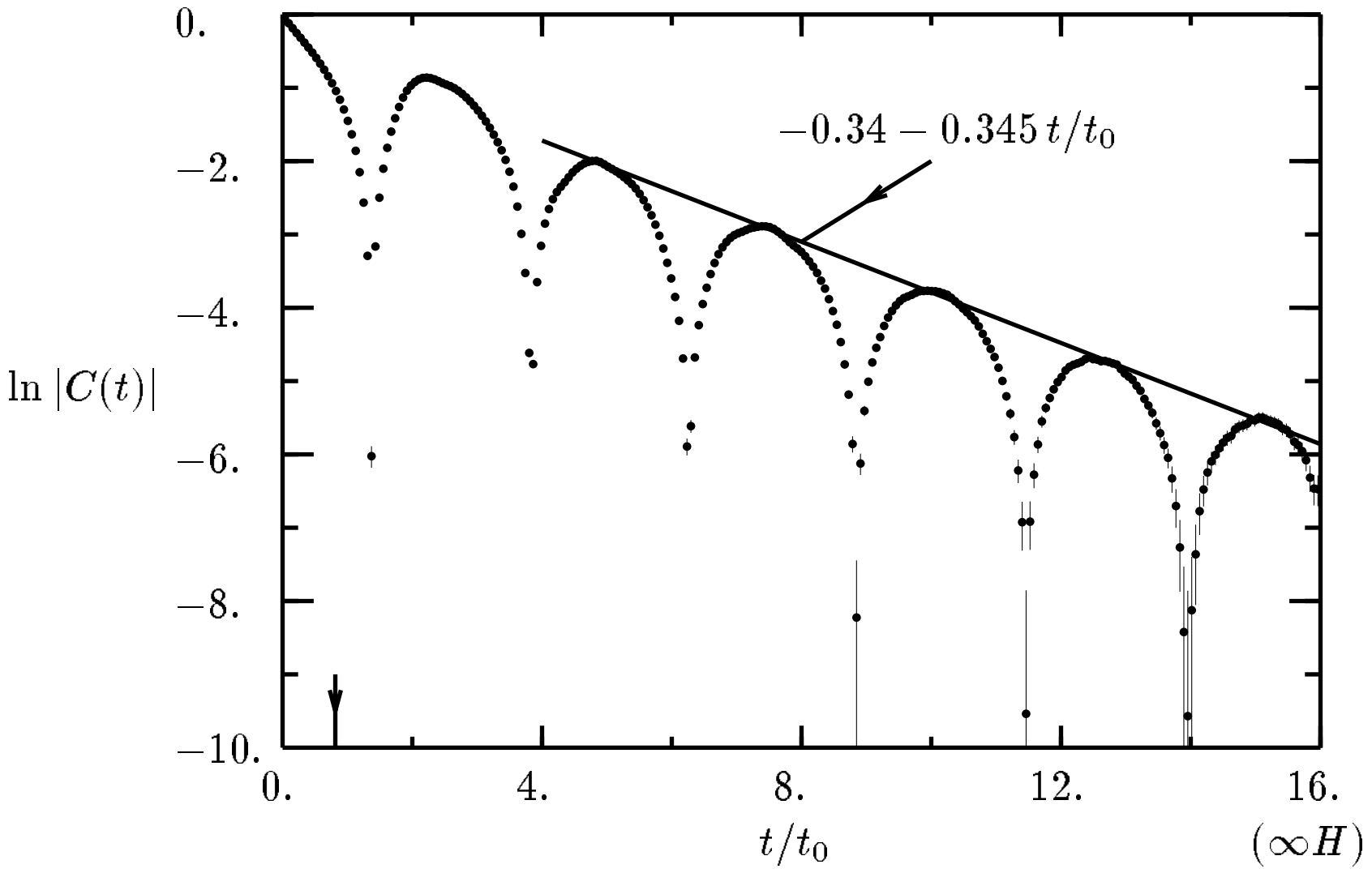

Fig. 3

*Logarithm of the velocity-velocity correlation function absolute value,* $\ln|C(t)|$*, versus* $t/t_0$ *for the* $\infty H$ *case. The solid line is the best linear fit of the last* 5 *local maxima of* $|C(t)|$*. The fit starts at twice the Lyapunov time scale (see comments at the end of* §2*).*

We also tried a three parameters best fit with the stretched exponential fit for the maxima getting an exponent 0.93 with error bars which does not include the unit value. Of course the fit is somewhat better and this is not surprising (clearly had we allowed for two more free parameters we would have obtained a perfect fit, see remarks in §1).

b): The 0H case:

We performed an analysis similar to the above. In **Fig.4** we see that $|C(t)|$ has more structure than in the $\infty$H case due to the more complex structure of the unit cell of the lattice of obstacles, (consisting of two sublattices with circles with different radius).

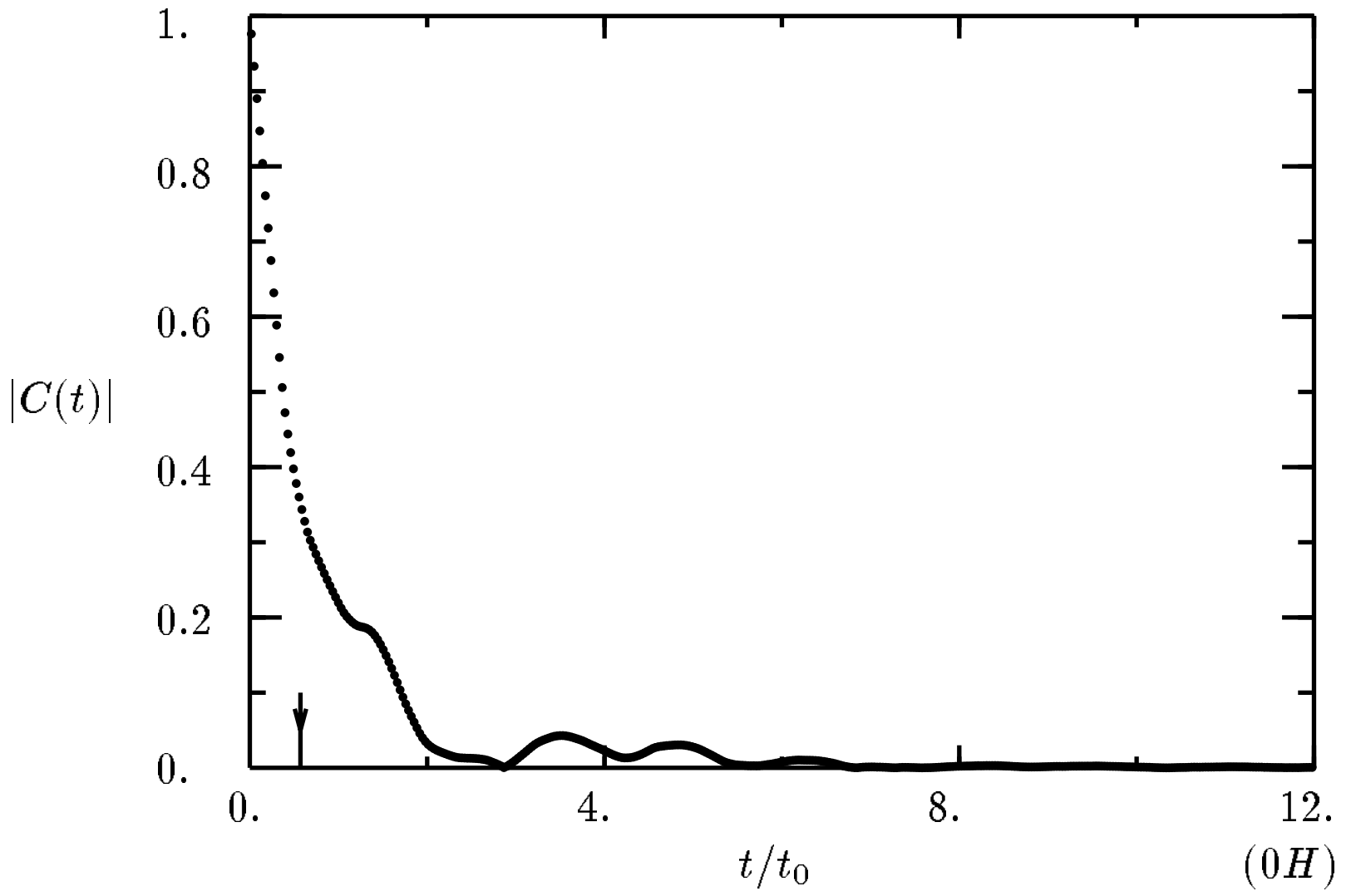

Fig. 4

*Absolute value of* $|C(t)|$*, the velocity-velocity correlation function, versus* $t/t_0$ *for the* $0H$ *case.*

The maxima of $|C(t)|$ have an apparently non trivial structure and this suggest a $e^{-at/t_0}f(wt)$ form of the curve with $f$ $2\pi$-periodic but more complicated in structure. This can be tested by a (rather severe) two parameter fit and the maxima should lie on parallel straight lines of the form $at_i/t_0 + b_i$ in a logarithmic plot.

When we plot $\ln|C(t)|$ versus $t$ (see **Fig. 5**), we see that there is in fact such a regular behaviour: there are two well defined similar cells with three maxima in each one. We get the fits for the two pairs of contiguous maxima: $-0.9(\pm0.3) - 0.54(\pm0.05)t/t_0$ and $-1.2(\pm0.1) - 0.54(\pm0.04)t/t_0$. In both cases the decay rate for the correlation function is $1.8(\pm0.1)\,t_0$, apparently uncorrelated to the Lyapunov exponent. Note that the error bars, at the maxima points, become too large after $t/t_0 \approx 12$.

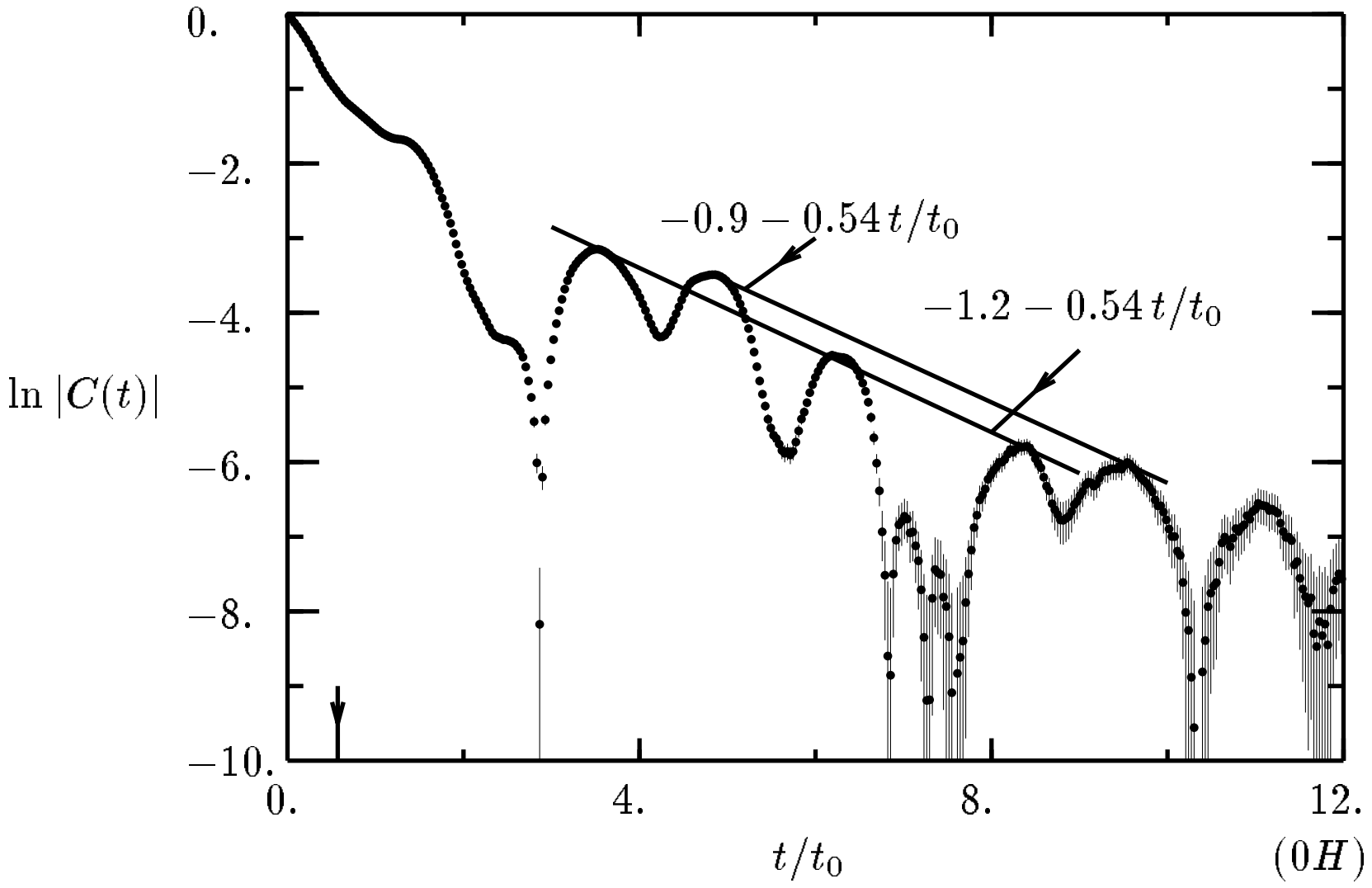

Fig. 5

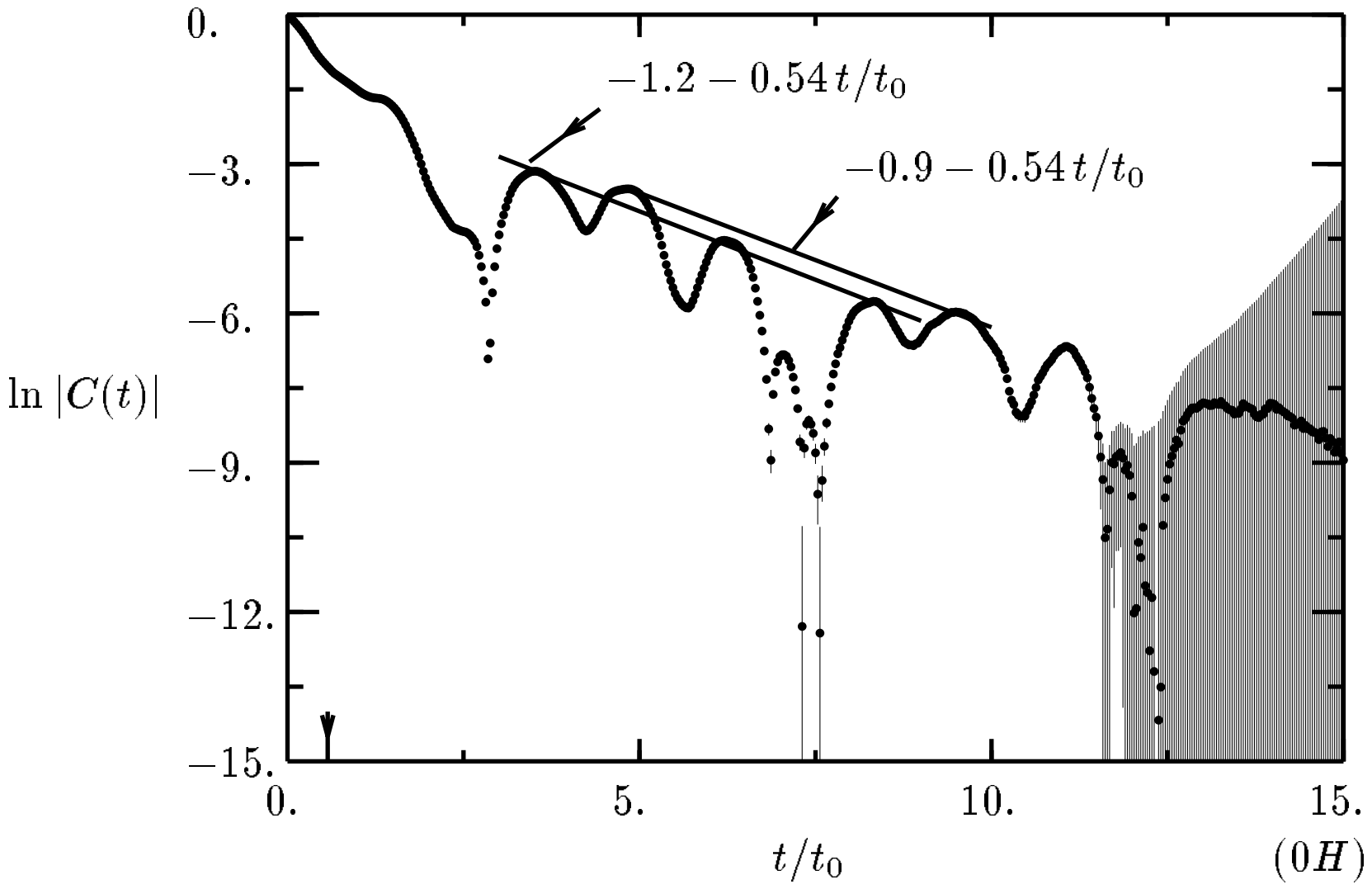

Fig. 6

*Logarithm of the velocity-velocity correlation function absolute value,* $\ln|C(t)|$*, versus* $t/t_0$ *for the* $0H$ *case with* $1 \times 10^7$ *trajectories (Fig. 5) and with* $2.8 \times 10^8$ *trajectories (Fig. 6).*

This experiment is very time consuming: we performed it with great care over a much longer time span than the others. The reason is that an experiment over a time scale comparable (computerwise) to the others would have yielded **Fig. 5**, which would have been quite inconclusive because of the error bars becoming too large after $9t_0$ (with the corresponding data *not* used in the fit). On the other hand the large number of relative maxima promised a very accurate test of our exponential decay assumption if the calculation could be pushed to the time $12t_0$, where in Fig. 5 the linear law looks possible only because of the large error bars. The longer time experiment results allows us to reduce error bars considerably up to $t \simeq 12t_0$. We reported the results also for $t$ larger that $12t_0$, when the error bars become visible just to give an idea of how difficult it could be to improve the analysis (which in the present form took three months of CPU on an IBM RISC-6000, just for Fig. 6).

Although the more refined experiment confirms and improves all the data of Fig. 5, the last clear maximum in Fig. 6 seems to be much higher than it should be in order to fit on the lower line. This might be due to corrections to the periodicity hypothesis or to actual stretching of the decay; or our definitions of errors are too generous (because, see Appendix, they (must) contain some elements of arbitrariness, as in (A1.6), which in such extreme data may require further analysis). This is a point that imposes further care, as we could not resolve it.

c): The Diamond case:

In **Fig.7** we show $|C(t)|$ in this case. Its structure is similar to the $\infty$H case: oscillations with decreasing amplitude with period $\tau = 1.72(\pm 0.03)t_0$. In this case we have very clean (errorwise) data up to times of order $13\,t/t_0$ as it can be seen more clearly from the following logarithmic plot.

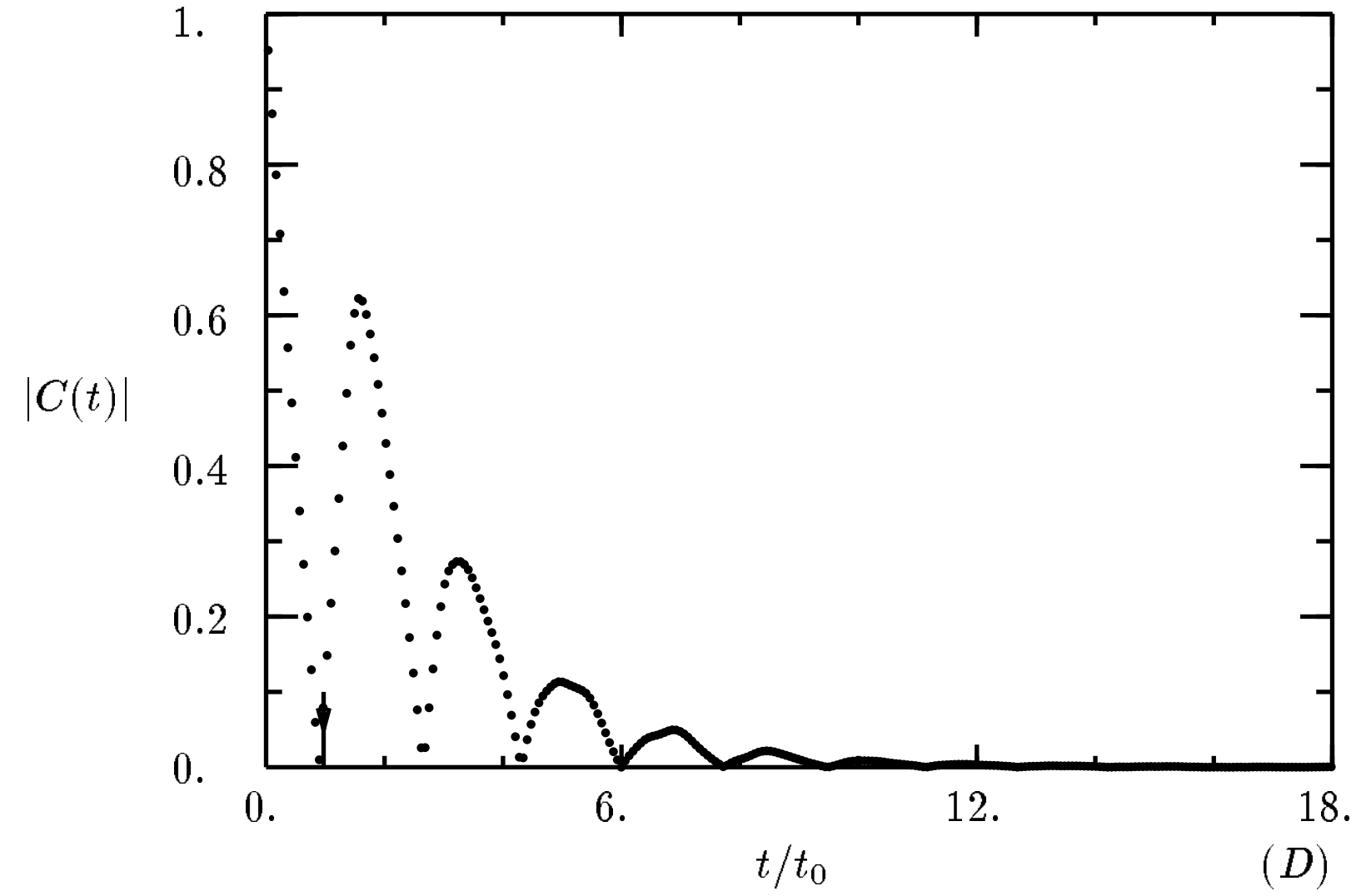

Fig. 7

*Absolute value of $|C(t)|$, the velocity-velocity correlation function, versus $t/t_0$ for the D case.*

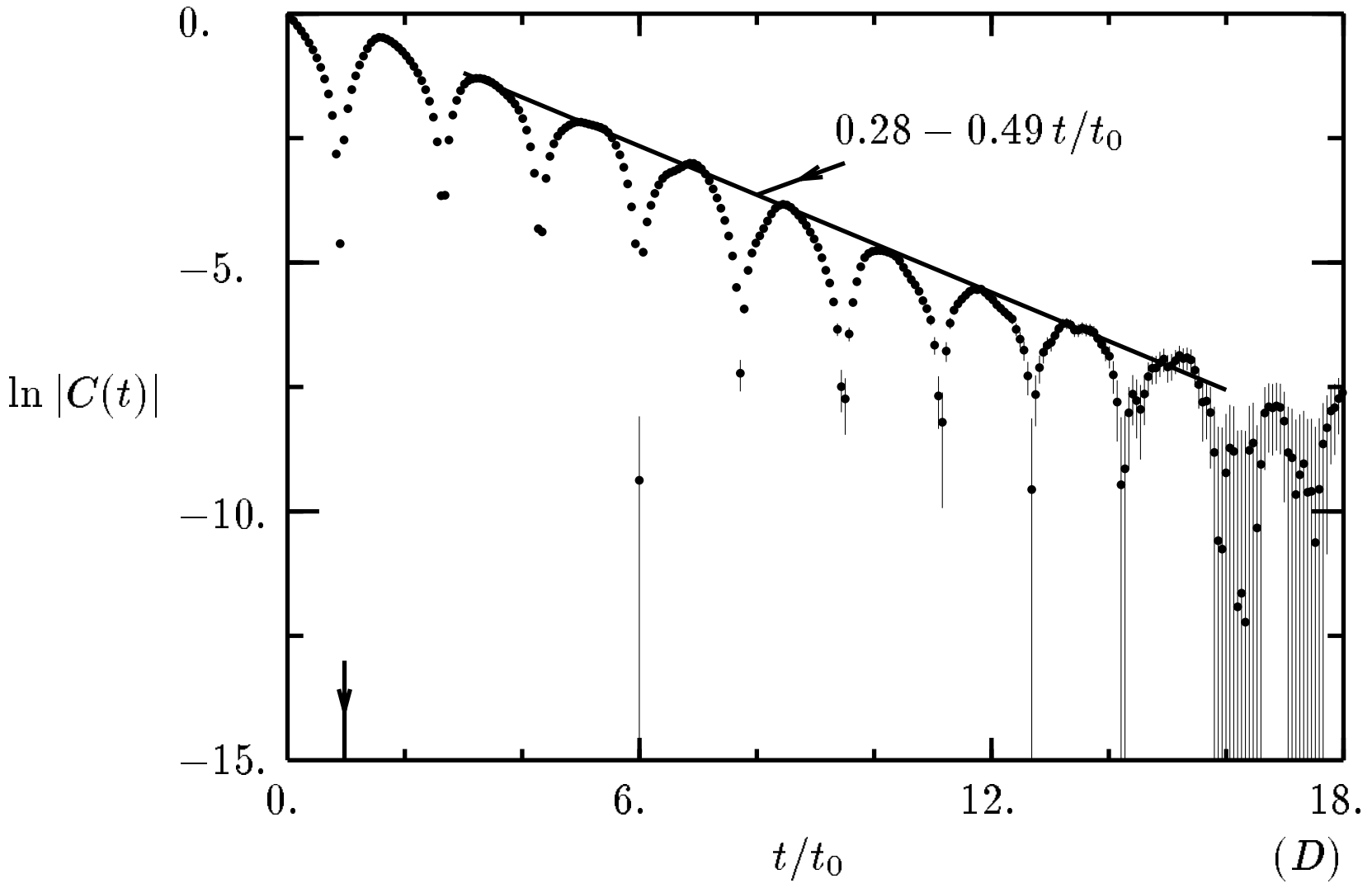

Fig. 8

*Logarithm of the velocity-velocity correlation function absolute value, $\ln|C(t)|$, versus $t/t_0$ for the D case. The solid line is the best linear fit of the last 8 local maxima of $|C(t)|$.*

The plot of $\ln|C(t)|$ versus $t$ shows again the pure exponential behaviour of the last 8 maxima amplitudes that we fit by the equation (**Fig.8**): $0.28(\pm 0.06) - 0.49(\pm 0.01)t/t_0$. The decay rate for the correlation function is now $2.04(\pm 0.04)t_0$, again apparently unrelated to the Lyapunov exponent. As in the ∞H case, if we try to fit a stretched exponential, we get an exponent 0.995 with an error which does not include the value 1 (see comments in §1 and to Fig. 4 above).

Comments:

There are not too many results in the literature for this autocorrelation function. In the ∞H case [BlD] get, from a computer simulation with a single trajectory with $2 \cdot 10^7$ collisions, that: $|C(t)| \sim e^{-at}$ for $t < t_0$, and $|C(t)| \sim e^{-bt^{0.7}}$ for $t_0 < t < 43t_0$. and, furthermore, referring to the decay, found in [FM]: $|C(t)| \sim ct^{-1}$ when $t \approx 110. t_0$ they remark that it is independent of $R$ when $R < 2^{-3/2}$. They also give arguments for a pure exponential decay when the system dimension goes to infinity. There is no reference to the rich structure of $C(t)$, clear in [FM], in the analysis of the decay.

The paper [FM] also contains, as mentioned in the introduction a large number of important results, that our experiments confirm; we mention: 1) They found for an "almost" triangular diamond a behavior similar to ours in Fig. 8 and they said explicitly that: "the velocity autocorrelation function appears to be an exponential damped periodic function". 2) They showed for the 0H case a rich structure for maxima and minina less ordered but similar to ours in Fig. 5 and 6. They say explicitly that "the curve appears to be bounded by an exponential envelope". 3) They did not show for the ∞H case the regular decay as we show in Fig. 3 for short times. But they show a rough consistency of a power law decay as $1/t$ for long times.

*§4 Collision velocity autocorrelation:* $C_c(n) = \langle v_{cx}(0)v_{cx}(n)\rangle_c$,

The results of this section are a byproduct of the numerical results obtained by performing the experiments described in the previous section with the exception of the ∞H case. Although we do not think that they are particularly significant for the reasons explained in §1, we report them as their study illustrates the problems on the data analysis on the collision systems stressed in §1.

a): The ∞H ($3d$) case:

We computed the velocity autocorrelation function corresponding to the hit number $n$. For large $n$ the noise due to statistical errors is important. Since the computer simulation was performed for a fixed time $t_m = 4.$ (in absolute units), each trajectory may have different number $N(n)$ of collisions for each $n$ in time $t_m$; the number of data $N(n)$ that we averaged to get $C_c(n)$, depends on $n$ and so does its statistical error (because for large $n$ there are large fluctuations in the number of the sampled trajectories that actually experienced $n$ collisions, see also §7). We find that for $n > 16$ collisions the errors (which, as always, have two natures: statistical and dynamical) become too large to permit setting up a data analysis.

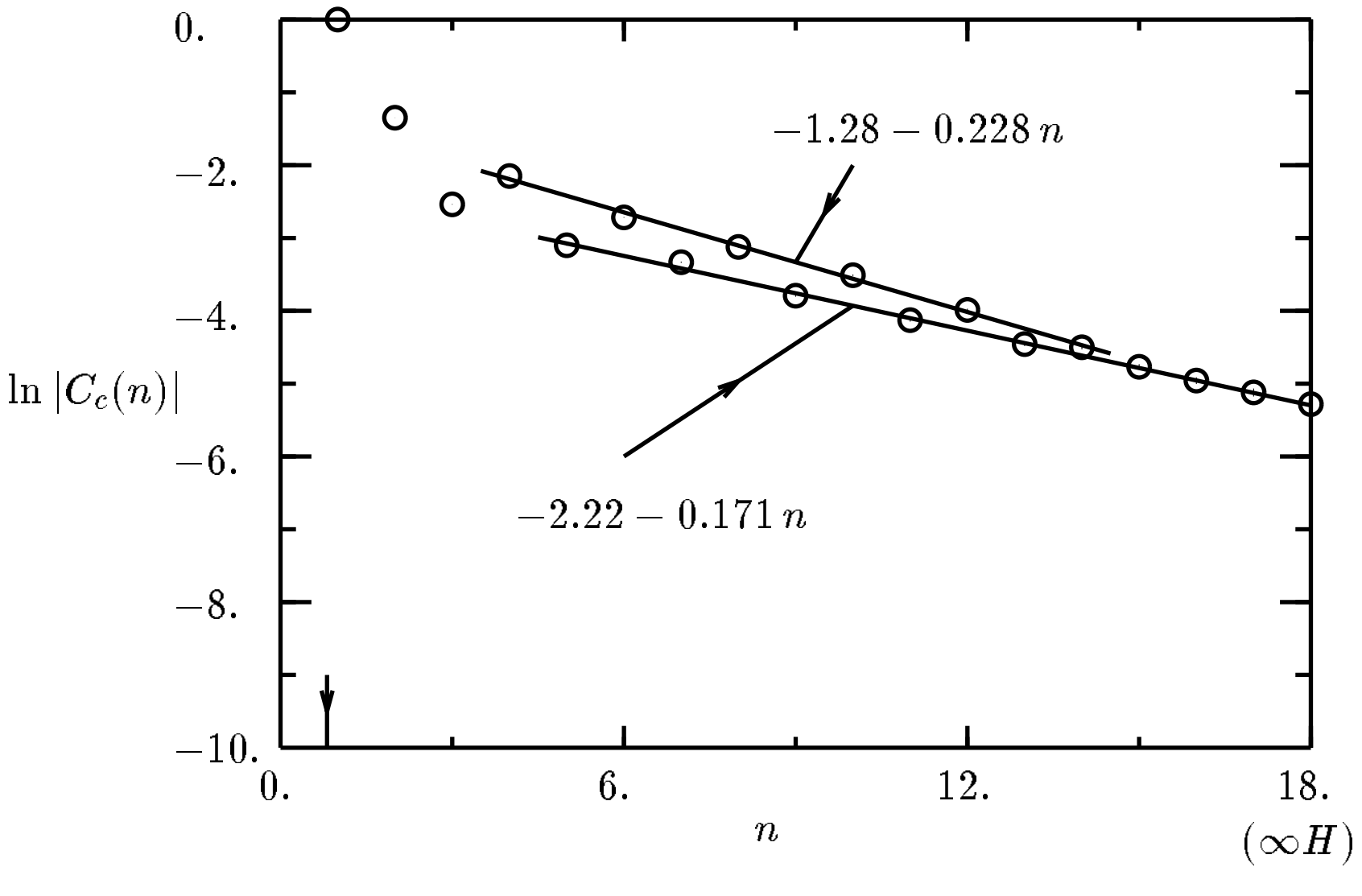

Fig. 9

*$\ln|C_c(n)|$ versus $n$ in the ∞H ($3d$) case. The straight lines are the best linear fits for the data points crossed by the lines.*

The **Fig.9** shows a $\ln|C_c(n)|$ versus $n$ plot. We see that there is a regular behaviour for the even and odd $n$ values separately. The two sets of values may be separately fitted by pure exponential laws as it is shown in the figure. The fits are: $-1.28(\pm 0.03) - 0.228(\pm 0.004)\,n$ and $-2.22(\pm 0.06) - 0.171(\pm 0.007)\,n$ for the upper and lower lines respectively. The decay rate for each set is $4.28(\pm 0.08)$ and $5.8(\pm 0.2)$ respectively. Also it seems that one of the sets merges in the other when $n > 15$.

The latter singular behavior may be perhaps understood by the discrete nature of $C_c(n)$. For instance, if the continuous time function had the analytical form: $C(t) = e^{-at}f(wt/t_0)$, being $f$ a periodic function with period $w^{-1}t_0$, (see §3a) then it would seem natural to expect that the collision one has the form: $C_c(n) = e^{-a'n}\bar{f}(w'n)$. Assuming that $w'$ is near a "resonance", say $w_0 = h/k$, with $h, k$ integers, and expanding the function $\bar{f}$ around $w_0$ we get: $\ln|C_c(n)| = -a'n + \ln\bar{f}(w_0 n) + (w' - w_0)n\bar{f}'(w_0 n)/\bar{f}(w_0 n)$. Therefore, we can get $k$ straight lines as the function $\bar{f}'/\bar{f}$ has period $k$, at least if $(w' - w_0)n$ is small.

In [BlD] $C_c(n)$ is computed by using a unique trajectory with $2\cdot 10^7$ collisions (we computed $2\cdot 10^7$ trajectories with 30 collisions each (at least), *i.e.* $6\cdot 10^8$ colllisions). They fitted $C_c(n) = (-1)^n e^{-an^b}$ for $1 < n < 9$ with $b = 0.86 \pm 0.06$. They do not mention the two branches that we observe in $C_c(n)$. We do not understand well enough their results to be able to reproduce or interpret them.

b): The 0H case:

In **Fig.10** we show the $\ln|C_c(n)|$ versus $n$ behaviour. In this case there do not seem to exist different branches for $n$ even and odd as they appeared in $\infty$H (it could be due to the fact that the period of oscillations may be far from a resonant regime and therefore, in the observed scale, the phenomenon is not relevant enough to be appreciated). All points fit the line: $1.1(\pm 0.3) - 1.2(\pm 0.1)\,n$ (see Fig.10).

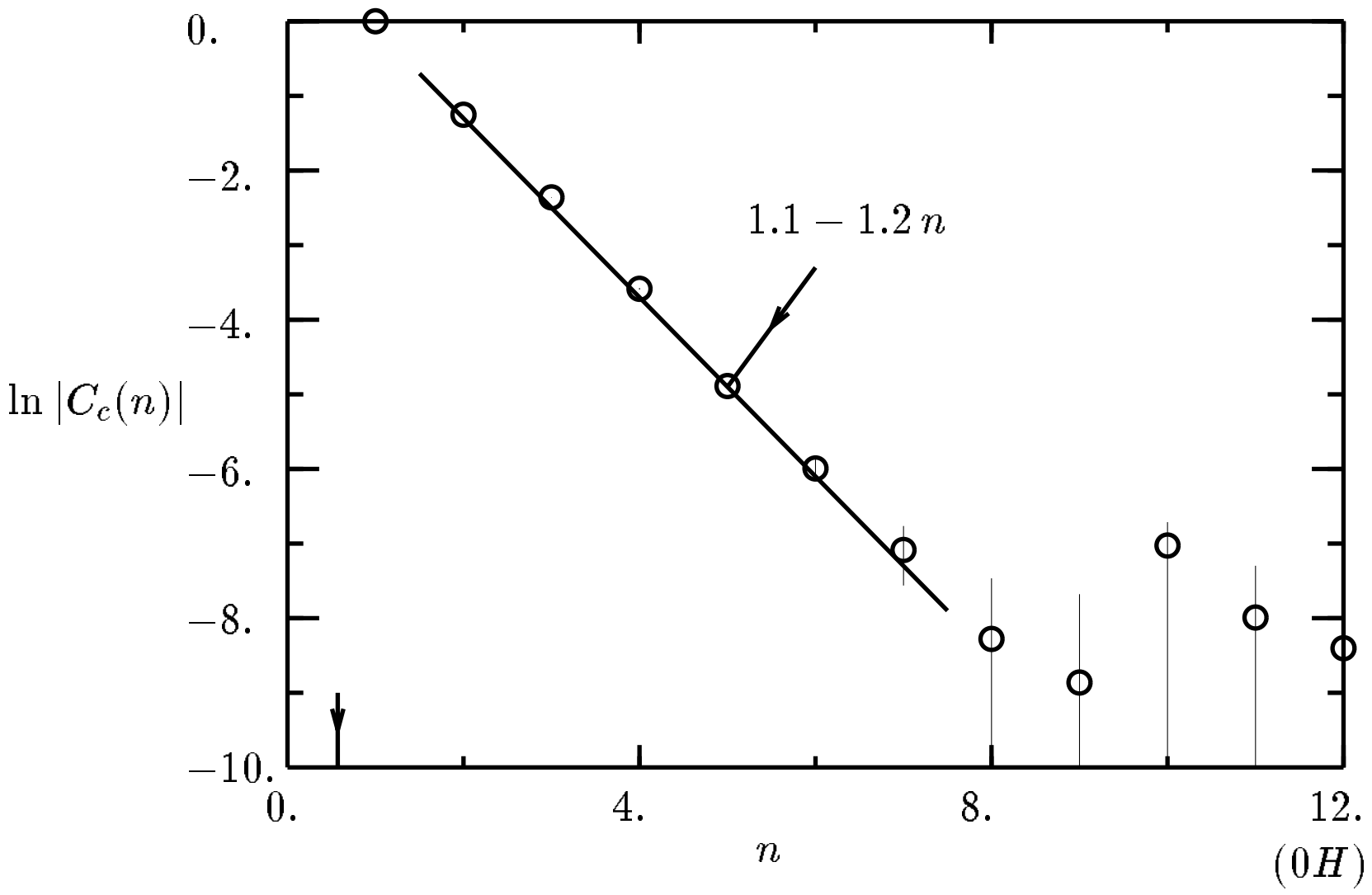

Fig. 10

$\ln|C_c(n)|$ *versus n in the 0H case. The line is the best linear fit for the data points it touches.*

Note how in this case the interval that can be analyzed only goes up to $n \approx 10$ compared with $n \approx 20$ for the $\infty$H case. That is so because the mean free time in this case is about twice the one for the $\infty$H case (because of our choice of the geometrical parameters) and therefore the number of collisions with good statistics is reduced by a factor of two. Even though in Fig.10 it is not shown, the data follow the rule: $sign(C_c(n)) = (-1)^n$.

c): The $D$ case:

The results in this case are less clear. In **Fig.11** we show $\ln|C_c(n)|$ versus $n$. The data, compared with the previous cases, have fluctuating behavior. We get some linear fits by using different $n$ intervals with relative errors in the coefficients larger than 15%. For instance, the one showed in Fig.11 is: $-1.(\pm 2.) - 0.5(\pm 0.3)\,n$. We also get a *three parameters* stretched exponential fit (see comments in §1 and following Fig.2) starting the minimization process randomly (with the parameters values ranging between $-5.$ and $5.$)

we found: $4.0(\pm 0.9) - 4.1(\pm 0.4)n^{0.41(\pm 0.03)}$. In order to compare with the non stretched exponential law we fix the value of the stretched exponent to 0.41 and we fit the remaining two parameter function. We get: $4.(\pm 2.) - 4.(\pm 1.)n^{0.41}$. The ratio between the *goodness* (see Appendix) of the exponential and stretched exponential behaviors is: 1.1. Plotting in a $3d$ graph the error function $G$ which measures the accuracy of the fits (see Appendix), as a function of the three parameters of the stretched exponential fit, say $(\alpha_0, \alpha_1, \alpha_2)$, being $\alpha_2$ the stretched exponent, one finds that there is a path connecting the values $(4.0, -4.1, 0.41)$ and $(-1., -0.5, 1.)$ which is almost flat, slowly going up from $G = 0.4281$ to $G = 0.4688$. We conclude that the avalaible data are by far not good enough to distinguish stretched exponential laws with stretching parameter $\alpha_2$ anywhere between 0.4 and 1.

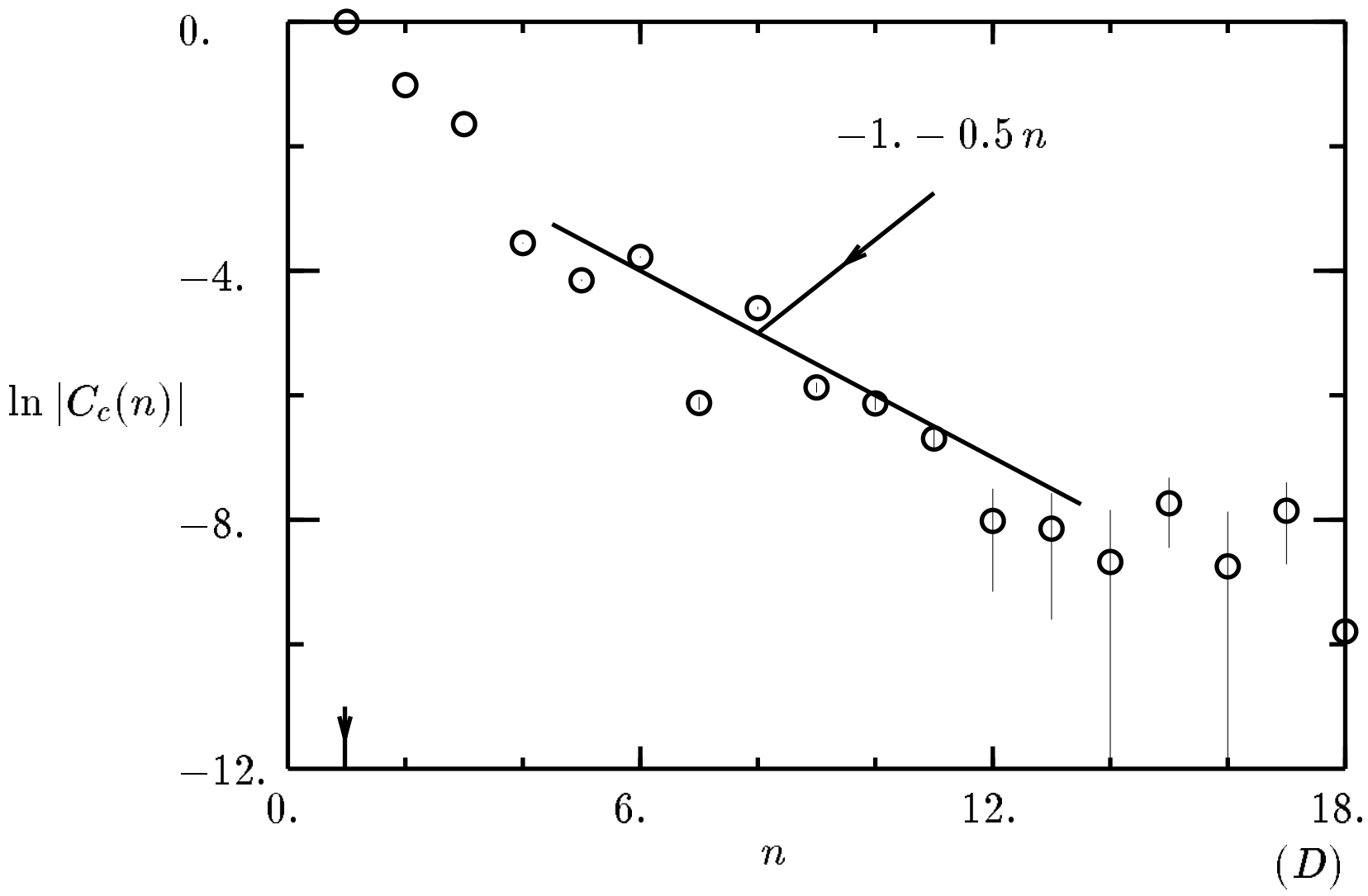

Fig. 11

*$\ln|C_c(n)|$ versus n in the D case. The line is the best linear fit for the data points lying in the same n-interval of the line.*

In [CCG] this case is studied by averaging $2.5 \times 10^5$ different trajectories and computing the quantity $r(n) = \langle \xi_A(0)\xi_A(n) \rangle$ where $\xi_A(n)$ is the characteristic function of set $A$ at time $n$ and $A$ is the set:

$$A = \{(x, y, v_x, v_y \,|\, x \in [0, 1/3], \, y \in [0, 1/2], \, \cos(1.1) \le v_x \le \cos(0.1) \\ \text{and } \sin(0.1) \le v_y \le \sin(1.1)\} \tag{4.1}$$

where the arguments of the trigonometric functions are measured in radians. The data are fit to the stretched exponential $r(n) = e^{-1.4n^{0.42}}$. Two main points from their computer simulation are not clear enough to us: there is an oscillatory character in their data which they eliminate by using a "suitable smoothing procedure", *i.e.* they seem to eliminate the "structure" of the data (a structure that might require further analysis as Fig.3 and Fig.8 illustrate). On the other hand they study trajectories with 70 collisions, and we cannot (as commented above) go beyond about 20 collisions still hoping to control the rounding error propagation and the statistical error which is too large to have good data. It would be necessary to know better the numerical method and data analysis actually used in the experiment [CCG] to clarify the above matters.

§5 *The transverse autocorrelation* $C^T(t) = \langle v_x(0)v_y(t) \rangle$.

We also computed the transverse velocity self correlation function for the $\infty$H ($3d$) case. In **Fig.12** and **Fig.13** we plotted its time behaviour: absolute value and the logarithm the absolute value versus time. The amplitude of oscillations is two order of magnitude smaller compared with the one in $C(t)$ and their period is equal. Again, the maxima follow an exponential decay behavior. In Fig.13 we see again how the maxima decay in a regular fashion and the linear fit of the last four maxima is: $-2.27(\pm 0.07) - 0.37(\pm 0.02)t/t_0$. The decay rate is equal to the one of $C(t)$ within error bars, *i.e.* $2.70(\pm 0.15)t_0$.

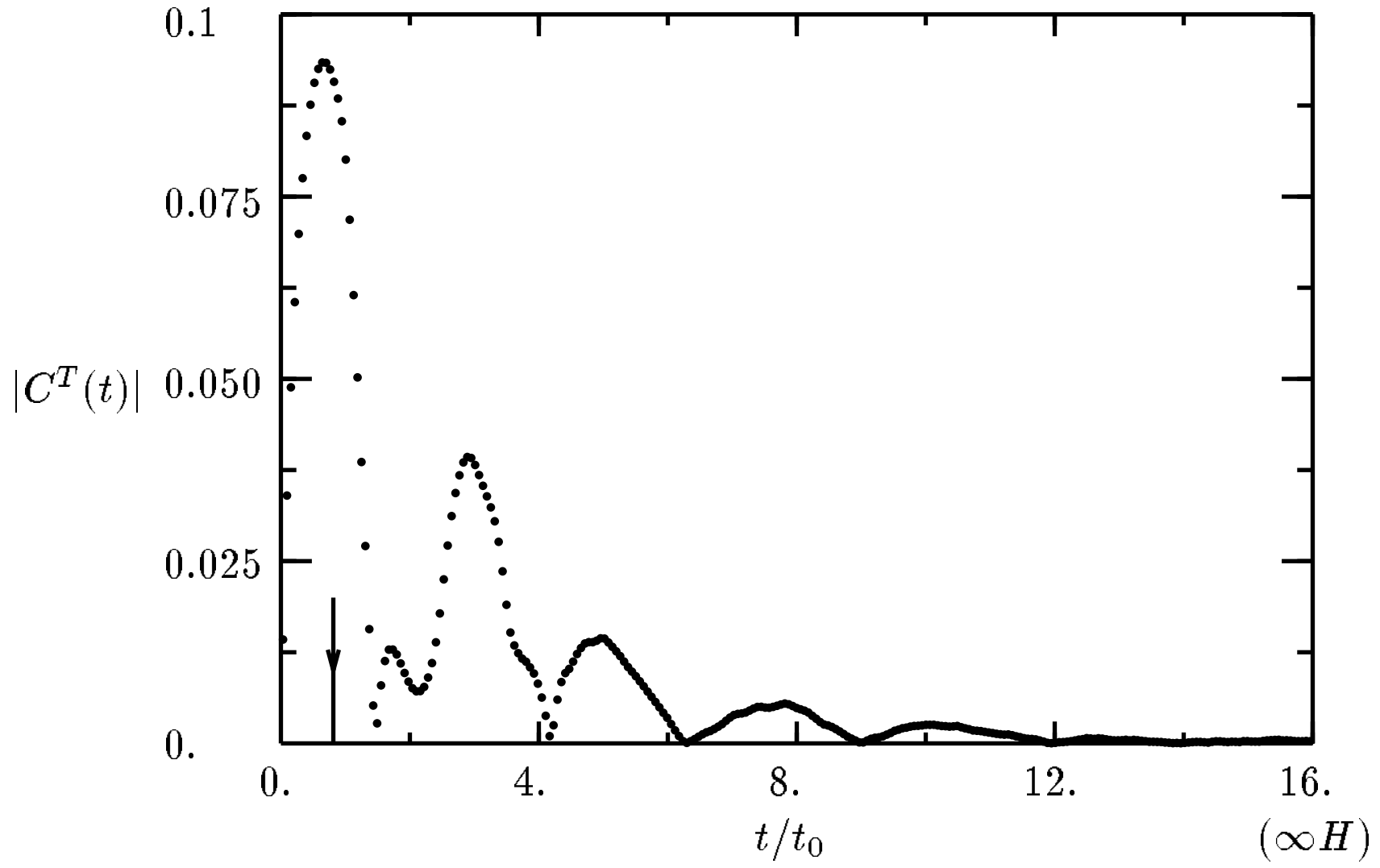

Fig. 12

*Absolute value of $|C^T(t)|$, the transverse correlation function (see eq. (1.4)), versus t for the $\infty H$ (3d) case.*

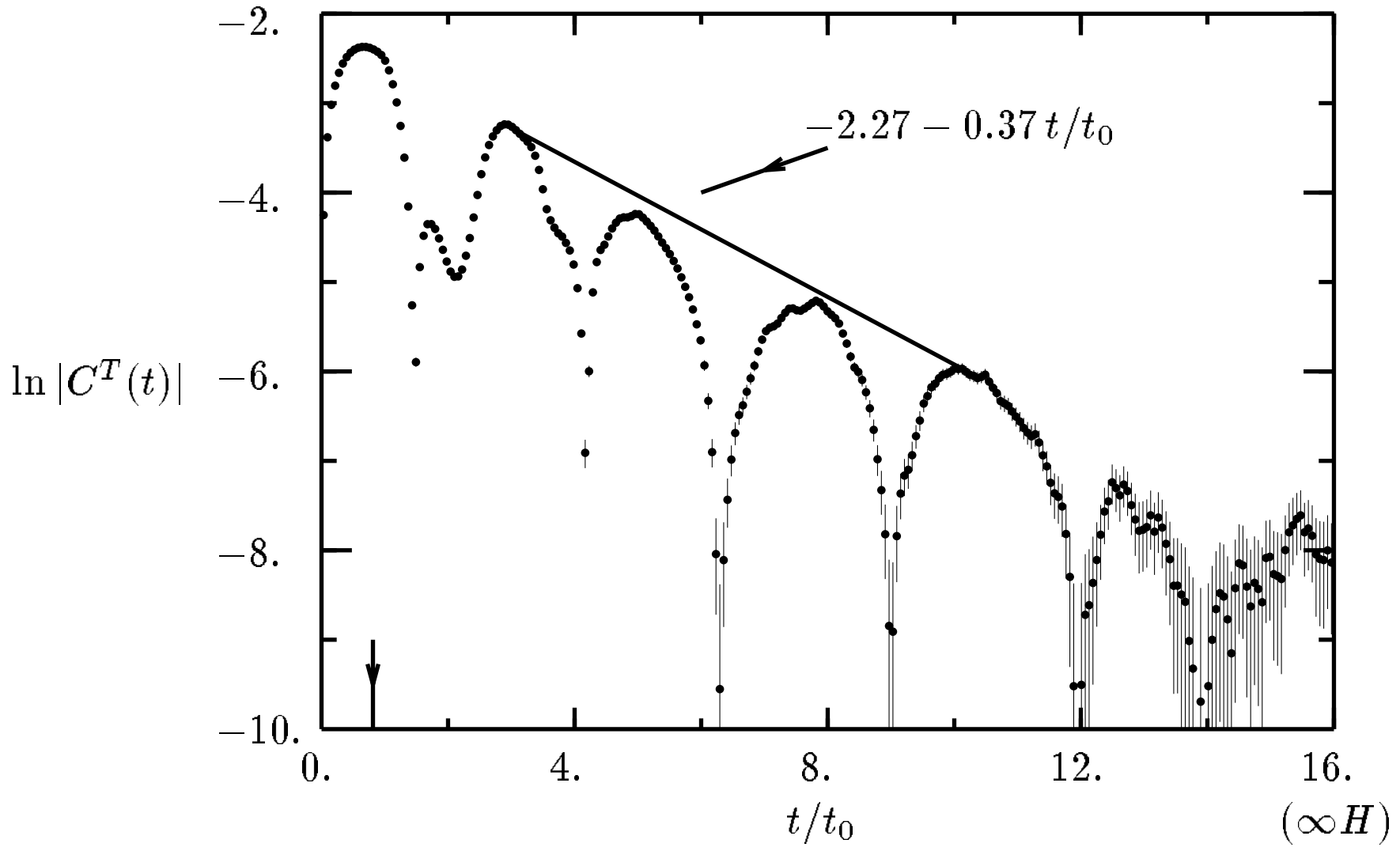

Fig. 13

*ln $|C^T(t)|$ versus t for the $\infty H$ (3d) case. The solid line is the best linear fit for the last four larger local maxima of $|C^T(t)|$.*

We have also computed the $C^T(t)$ in the 0H and $D$ cases, but we do not report the data (as no new information comes from them).

## §6 *The diffusion coefficients and* $s(t) = \langle x^2(t)\rangle$, $s_c(n) = \langle x^2(n)\rangle_c$ .

a) The $\infty$H ($2d$ and $3d$) case.

In **Fig.14** we show the mean square particle displacement, $s(t)$, versus time for the $2d$ and $3d$ systems. In both cases and for short times there is some structure due to the correlation with the initial condition (in particular we see the expected parabolic behavior coming from the free motion when the time is near zero). Notice also how the $2d$ system has a larger mean square displacement because its mean path to the first collision is larger. The latter explanation looks like inconsistent with the existence of unbounded trajectories for the $3d$ system. In fact, there exist trajectories with free paths larger than the time interval used ($i.e.$ $t_m$) but their number is so small that they, effectively, don't contribute to the averages. The asymptotic lines for the $2d$ and $3d$ cases are respectively: $0.069(\pm 0.04) + 0.0077(\pm 0.0003)t/t_0$ and $0.0369(\pm 0.0006) + 0.00758(\pm 0.00004)t/t_0$.

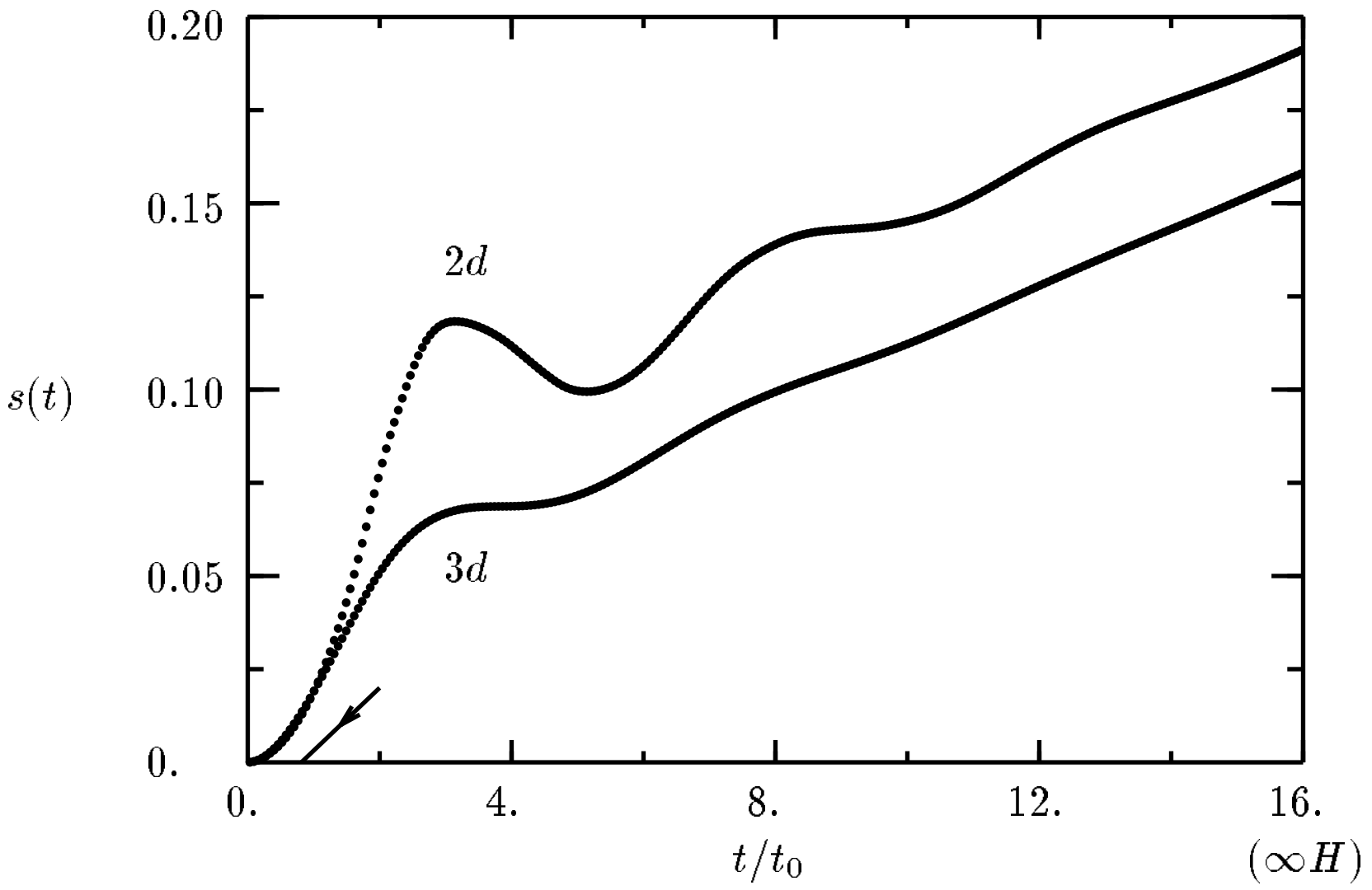

Fig. 14

*Mean square displacement, $s(t)$, versus time $t$, for the $\infty H$ ($3d$ and $2d$) case. The equations are the best linear fits for the $2d$ and $3d$ data which is larger than $14.t_0$*

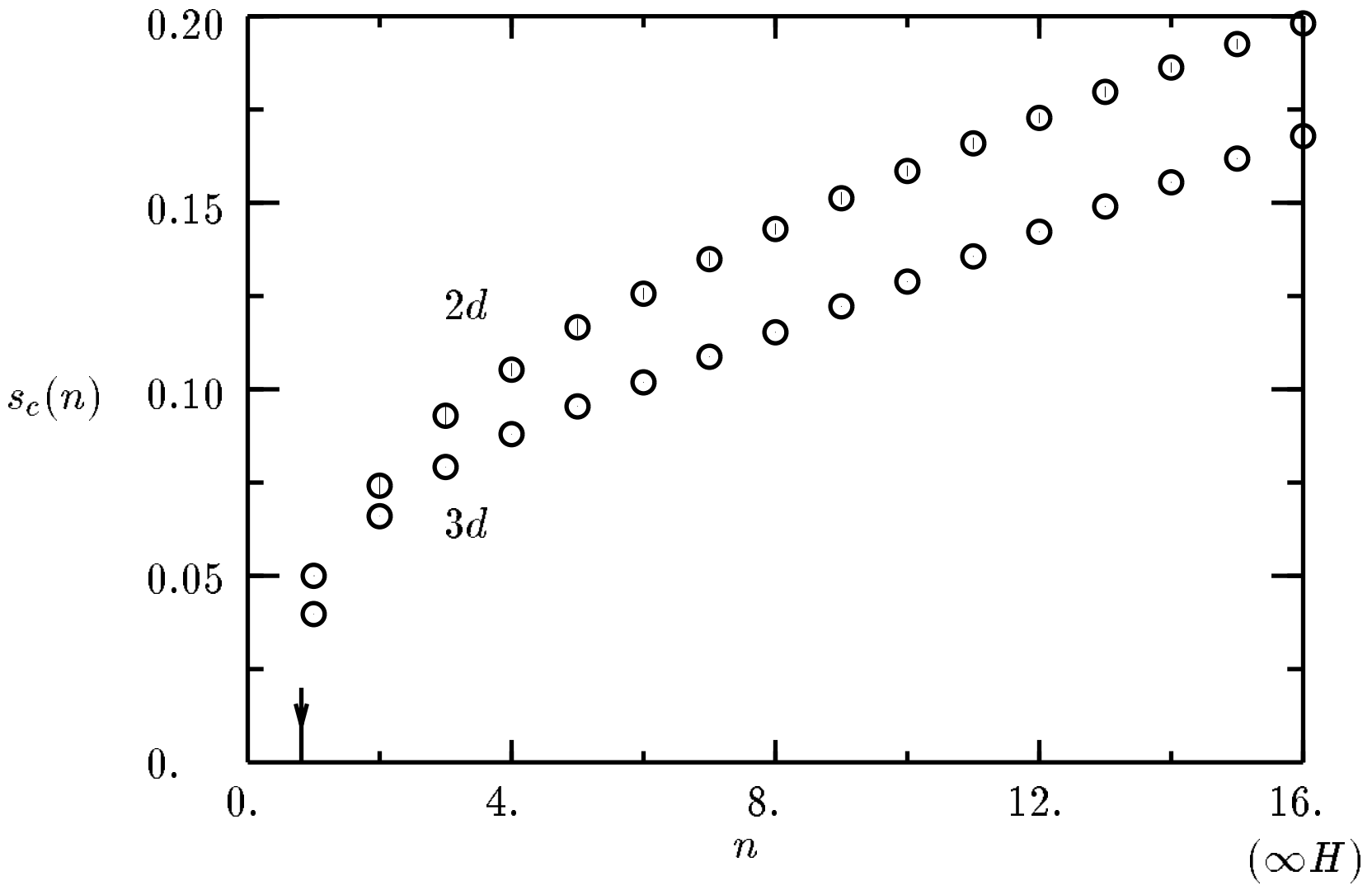

Fig. 15

*Mean square displacement, $s_c(n)$, versus the collision number, $n$, for the $\infty H$ $3d$ and $2d$ cases. The equations are the best linear fits for the $2d$ and $3d$ data which is between 8 and 16.*

Recently [Bl] argued that for the $2d$ system, the mean square displacement has the superdiffusive behavior: $s(t) = a + D_B t \ln(t) + Dt$ when $t$ is large enough. We tried to fit this behavior to our data but we found that $D_B \simeq 0.0001 t_0^{-1}$ and $D \simeq 0.0075 t_0^{-1}$, where the error is in the fourth significative digit ($i.e.$ 100% !). That is, the time interval we are using is too small to detect clearly the asymptotic superdiffusive behavior. In fact, we are far from the asymptotic regime in order to compare with the theoretical results by [Bl]. The latter remark is corroborated when we get $D_B$ following its analytical expression given in [Bl]:

$$D_B = \frac{1}{\sqrt{2}t_0} D_B^0(\sqrt{2}R) \tag{6.1}$$

and:

$$D_B^0(x) \equiv \frac{1}{2\pi x}(1-2x)^2 \tag{6.2}$$

having converted the results to our units (with lattice spacing 1) and $R$ is the obstacle radius (in the same units). In our case it gives the value $D_B = 0.00014 t_0^{-1}$ which is of similar order of magnitude of our computer simulated value.

The **Fig.15** also shows $s_c(n)$ versus $n$. The asymptotic lines for the $2d$ and $3d$ cases are respectively: $0.097(\pm.007) + 0.0063(\pm.0004)n$ and $0.0720(\pm.0007) + 0.00591(\pm.00005)n$. If we fit the funtion $s_c(n) = a' + D'_B n \ln(n) + D'n + b'n^{-1}$ to the data we find the same result for the $a'$ and $D'$ parameters, being $D'_B$ zero within our error bars. Finally, the relation $D/D'$ gives: $1.22(\pm0.13)t_0$ and $1.28(\pm0.011)t_0$. The discrepancy, far beyond the mean square deviation, with respect to the expected result $s(t)/s_c(n) = t_0$ when $t, n$ go to $\infty$ ([Bl]) possibly indicates again that we are far from the asymptotic region.

b): The 0H case:

In **Fig.16** and **Fig.17** we show $s(t)$ and $s_c(n)$ respectively. We get in both cases asymptotic linear behaviours: $a + Dt/t_0$ with $a = -0.026(\pm0.003)$, $D = 0.1129(\pm0.0004)t_0^{-1}$ and $a' = 0.027(\pm0.001)$, $D' = 0.1072(\pm0.0002)$ and then $D/D' = 1.053(\pm0.006)t_0$. We see that the error bars does not include the expected exact result $D/D' = t_0$: we think that is because we are not yet in the asymptotic regime.

Note we do not fit the last three data points in Fig.17. This is so because the time interval used ($t_m = 4$) is relatively short and some particle trajectories do move long enough to experience more than 8 collisions (*i.e.* they are trajectories with long free paths) and therefore they are relevant for the average but they are not contributing to the final value of $s_c(n)$ (in other words, at absolute time $t_m = 4$ there are trajectories which have not yet experienced $n$ collisions; their number is negligible if $n \leq 9$ but starts becoming important if $n > 9$, see Fig. 21 below).

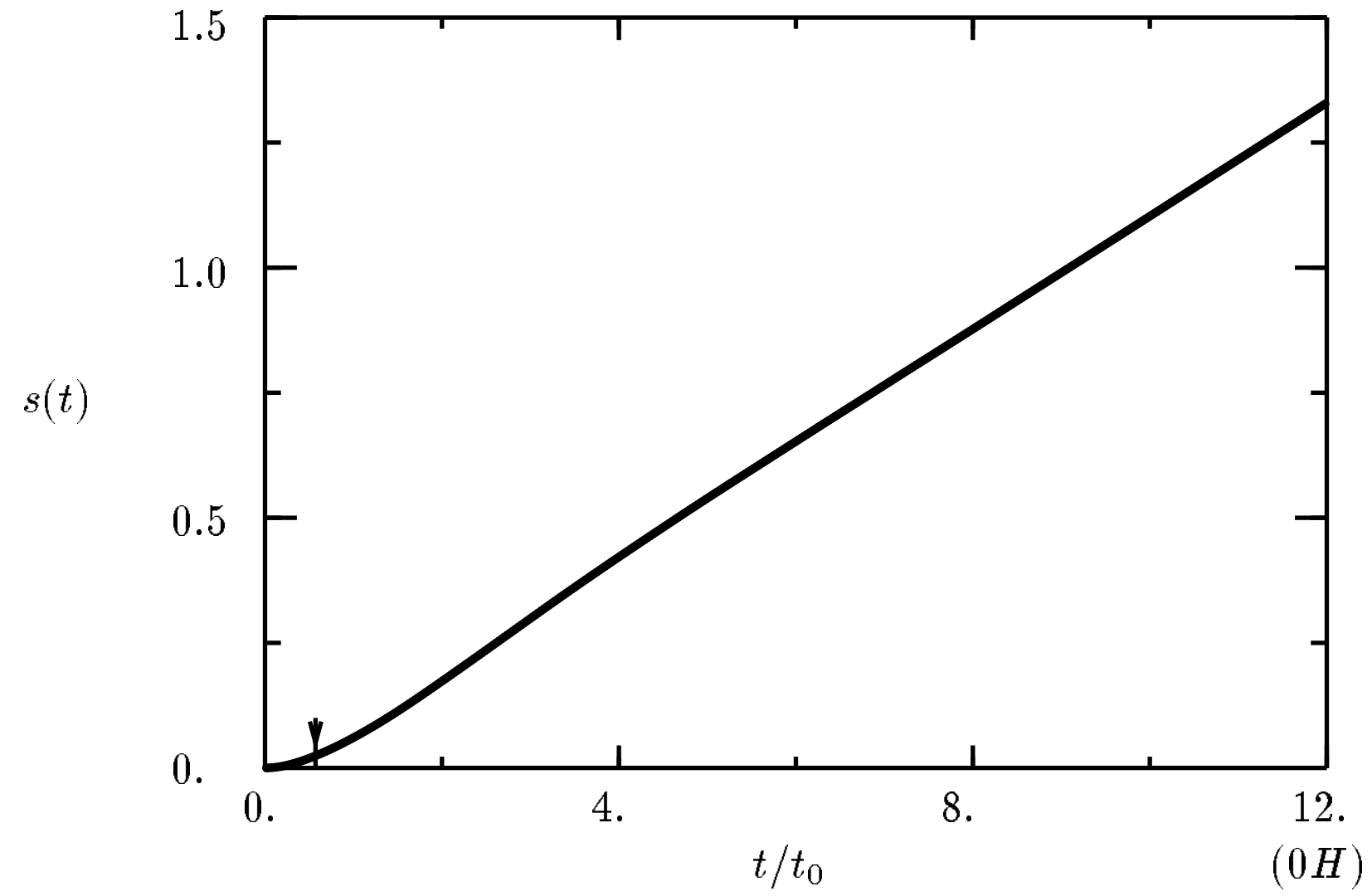

Fig. 16

*Mean square displacement, $s(t)$, versus time $t$, for the 0H case. The equation is the best linear fit for the data which is larger than $6\,t_0$*

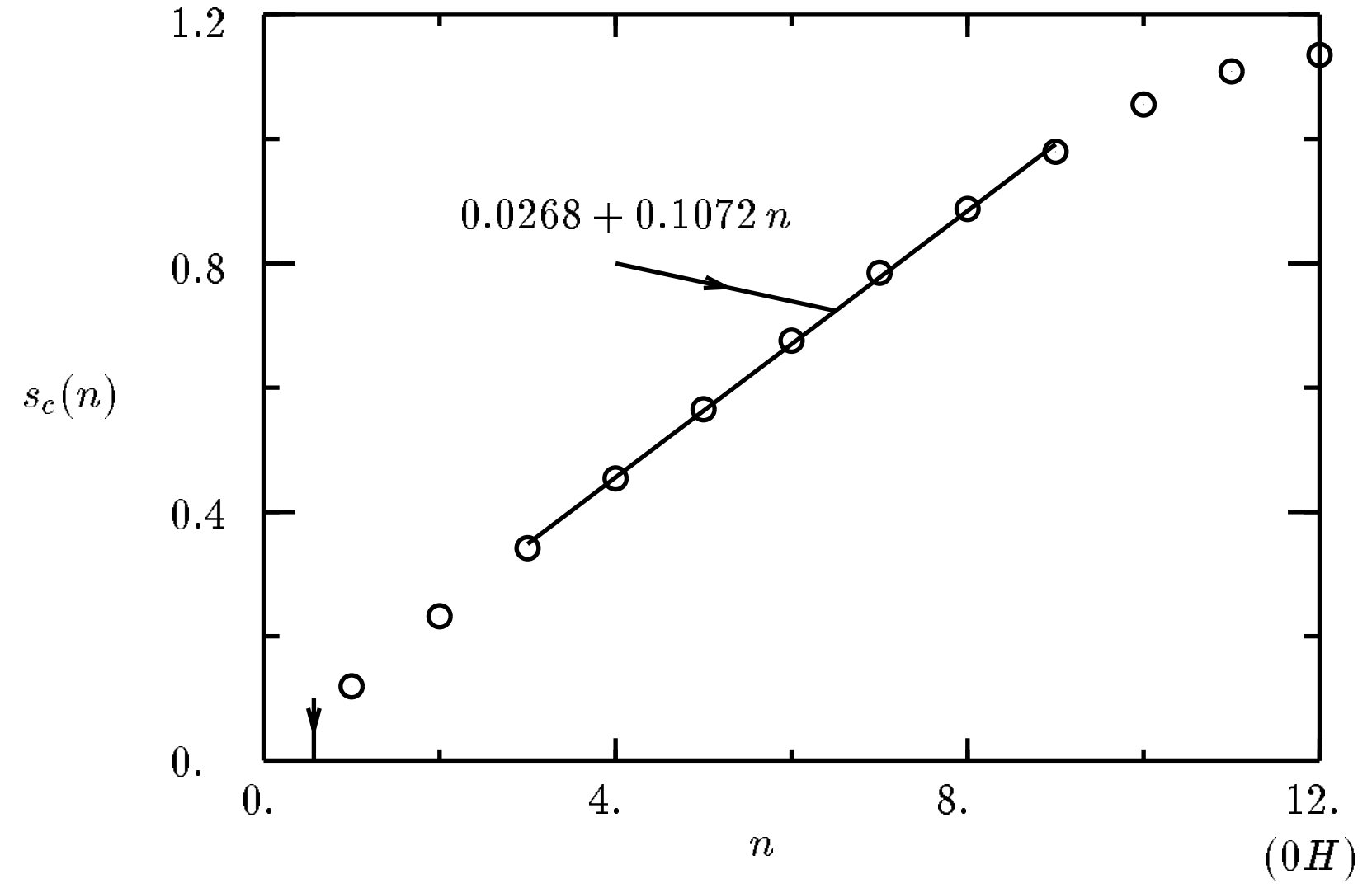

Fig. 17

*Mean square displacement, $s_c(n)$, versus the collision number, $n$, for the $0H$ case. The equation is the best linear fit for the data with $n$ is between 4 and 9.*

c): The $D$ case:

In **Fig.18** and **Fig.19** we show $s(t)$ and $s_c(n)$ respectively. We get the expected non-diffusive behavior: $s(t)$ and $s_c(n)$ tend to the limiting constant values 0.017 and 0.026 respectively. We can compare, to check the ergodicity, such numerical values with the ergodic averages:

$$s(+\infty) = \frac{\int d\vec{x} d\vartheta\, \vec{x}^2}{\int d\vec{x} d\vartheta} = 0.01727..., \qquad s_c(+\infty) = \frac{\int dsdc\, \vec{x}^2}{\int dsdc} = 0.0264... \tag{6.3}$$

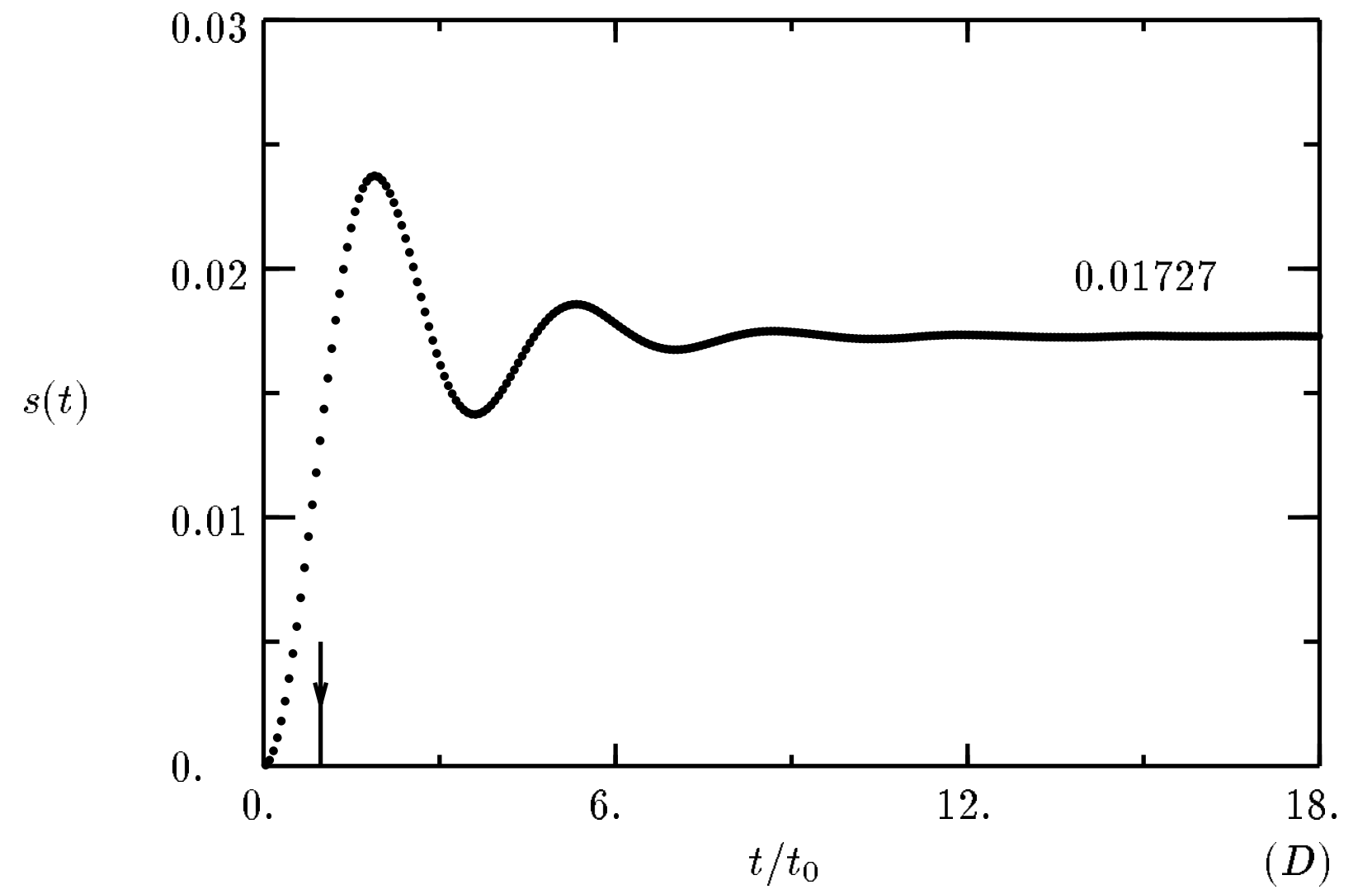

Fig. 18

*Mean square displacement, $s(t)$, versus time $t$, for the $D$ case. The number is the asymptotic exact value (see text).*

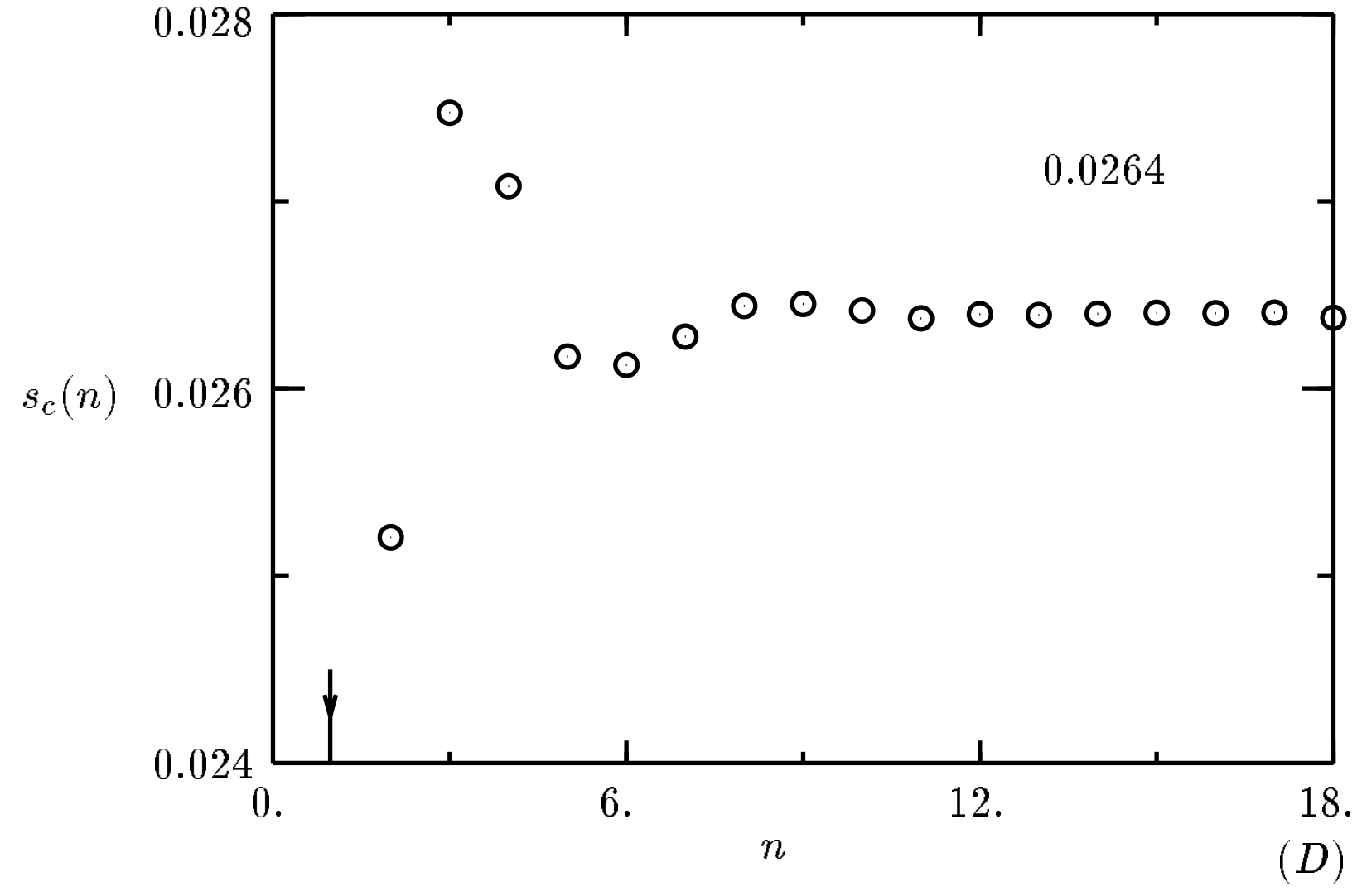

Fig. 19

*Mean square displacement, $s_c(n)$, versus the collision number, $n$, for the D case. The number is the asymptotic exact value (see text).*

This is a nice check of the ergodic theorem ([S]) and of course it could be repeated for many other averages and used as a test of how much one could expect to be in an asymptotic regime. Such a test would probably be necessary if one wanted to perform a more complete analysis of the $\infty$H and 0H cases. In the D case, as we see, there are no reasons to expect that the asymptotic regime has not yet been reached in the observed time intervals.

§7 *The collision number distribution: $d(n; t_m)$.*

We computed the probability distribution, $d(n; t_m)$, that the particle undergoes $n$ collisions in the time interval $[0, t_m]$ for $t_m = 1$, 1.5, 2, 2.5, 3, 3.5 and 4. In **Fig.20, 21** and **22** we show $d(n; t_m)$ versus $n$ for the $\infty$H ($3d$), 0H and $D$ cases respectively.

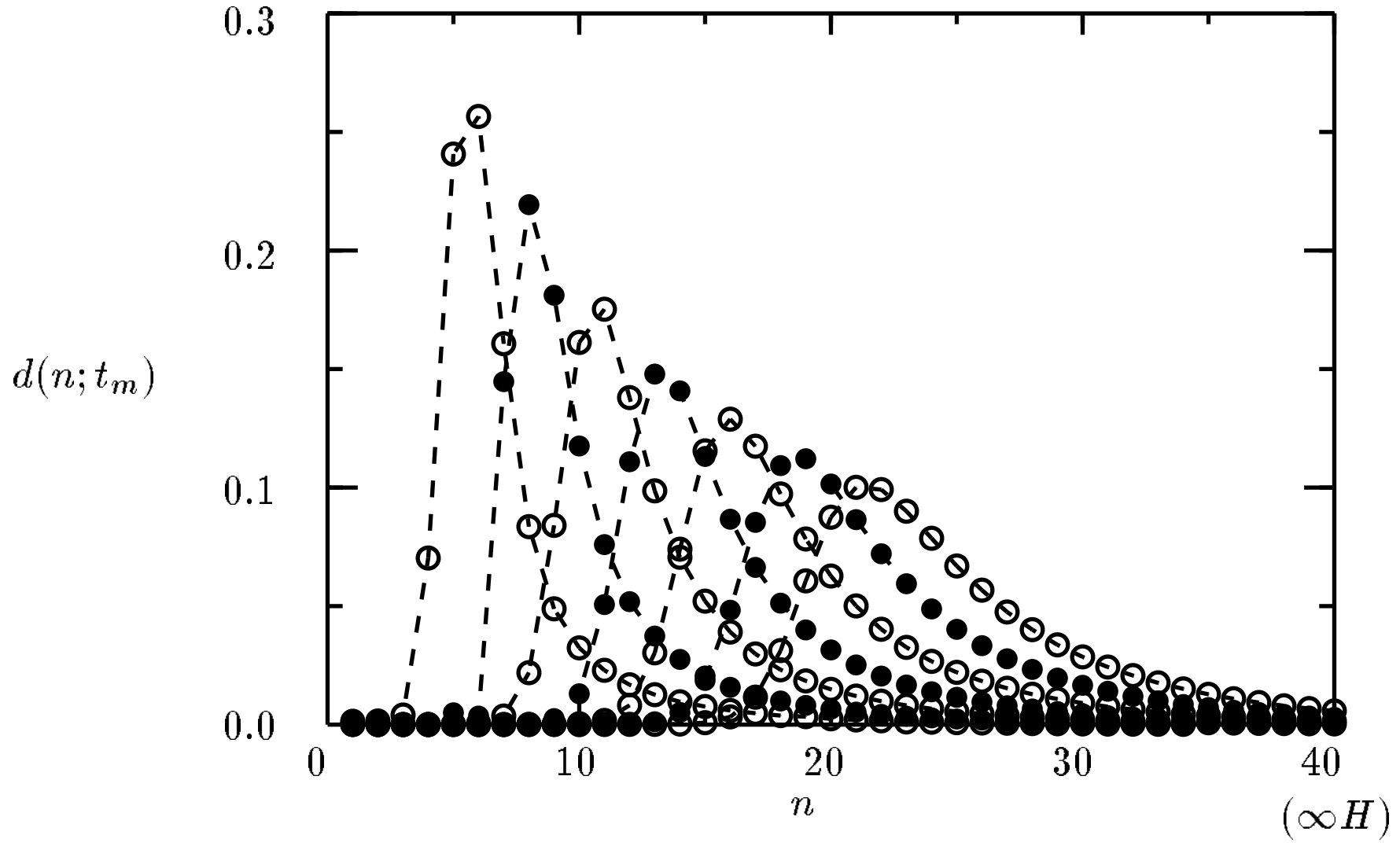

Fig. 20

*The distribution probability of the number of trajectories having $n$ collisions in a time interval $[0, t_m]$, $d(n; t_m)$, in the $\infty$H (3d) case for (from left to right) $t_m = 1.$, 1.5, 2., 2.5, 3., 3.5 and 4.; the dotted lines are visual guides.*

The aim of this analysis was to understand the reliability of the calculations on the ∞H case. We also performed the same calculation for the 0H and $D$ cases (see below); in the latter cases the computation was done in order to compare with the ∞H case. As a byproduct we could test in the three cases the natural hypothesis that a suitable rescaling leads to a gaussian distribution of the number of collisions in a given time.

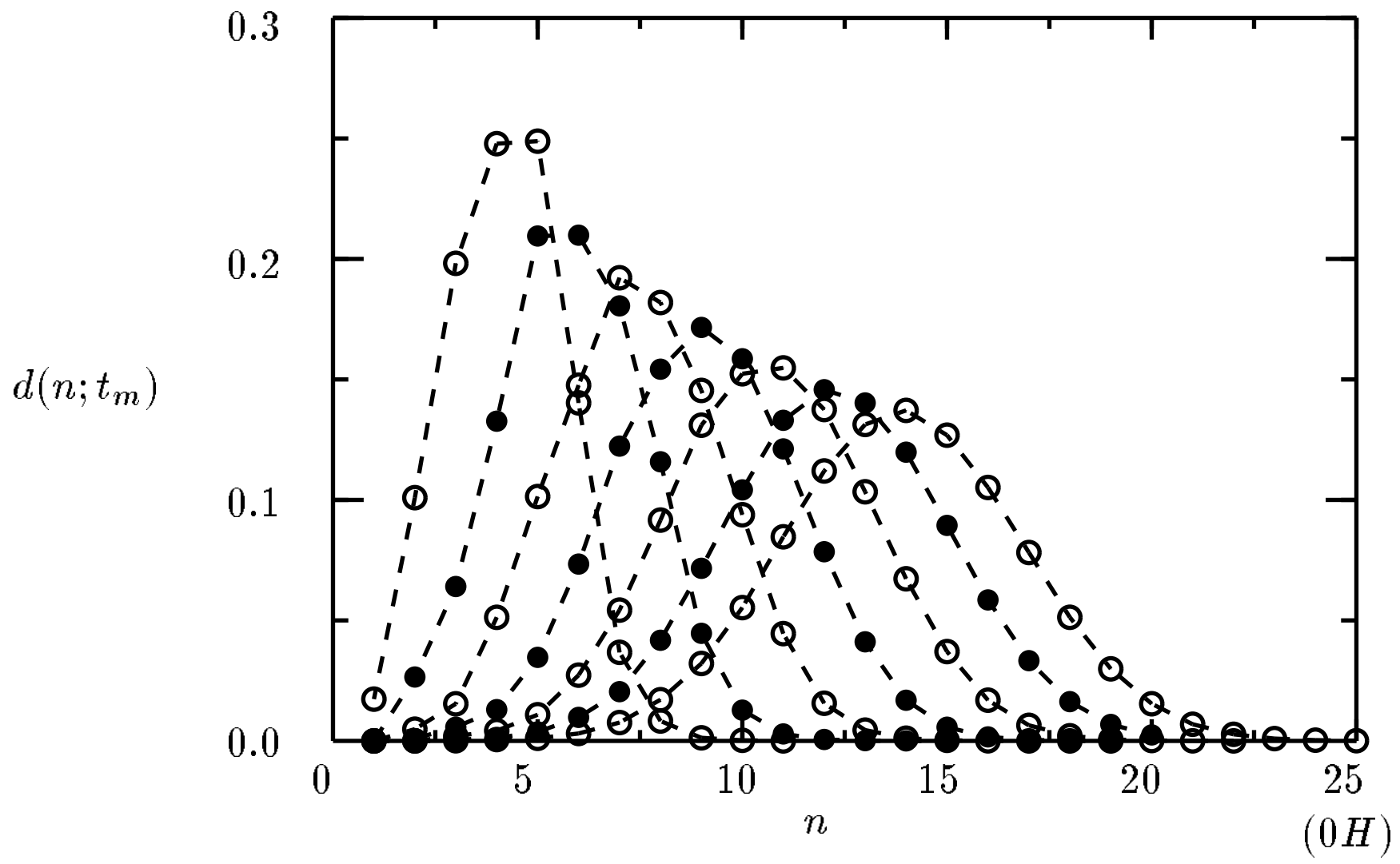

Fig. 21

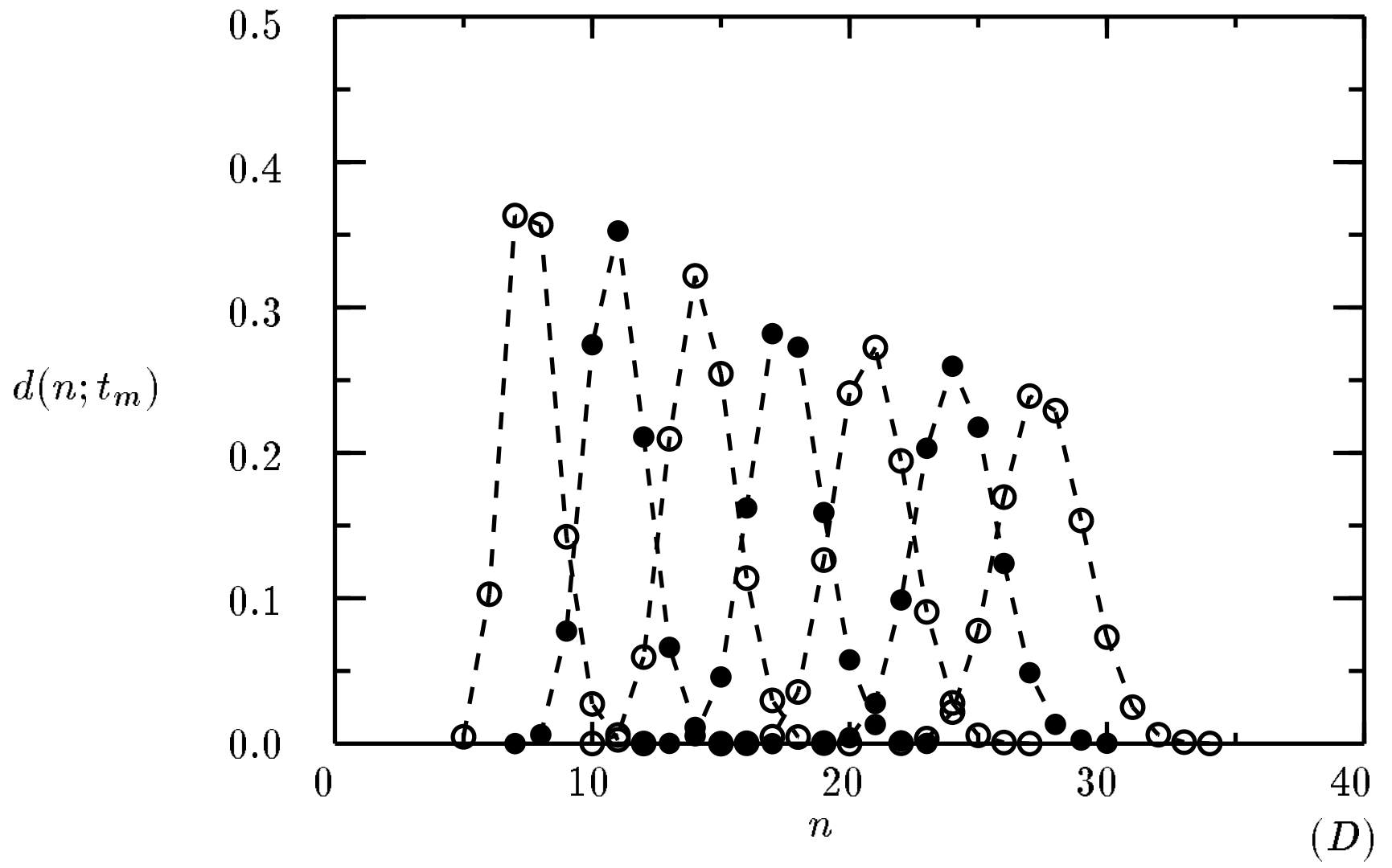

Fig. 22

*The distribution probability of the number of trajectories having $n$ collisions in a time interval $[0,t_m]$, $d(n;t_m)$, for different $t_m$ in the 0H (Fig. 21)and D (Fig.22) cases for (from left to right) $t_m = 1.$, 1.5, 2., 2.5, 3., 3.5 and 4..*

We see that, except in the ∞H ($3d$) case, the distributions look like gaussians and, in each case, their structure evolves with $t_m$ in a regular form. In particular in **Fig.23** we show the averaged number of collisions per trajectory, $\langle n\rangle_{t_m}$, for all three cases and different $t_m$. We see how the linear fit is almost perfect in all cases: $0.9998(\pm 0.0001) + 6.622(\pm 0.004)t_m$, $0.99836(\pm 0.00008) + 6.081(\pm 0.004)t_m$ and $1.0005(\pm 0.0004) + 3.228(\pm 0.002)t_m$ for the D, 0H and ∞H cases respectively. The inverse of the linear fit slope gives in all cases the mean free time (within errors) $t_0$ for each case respectively.

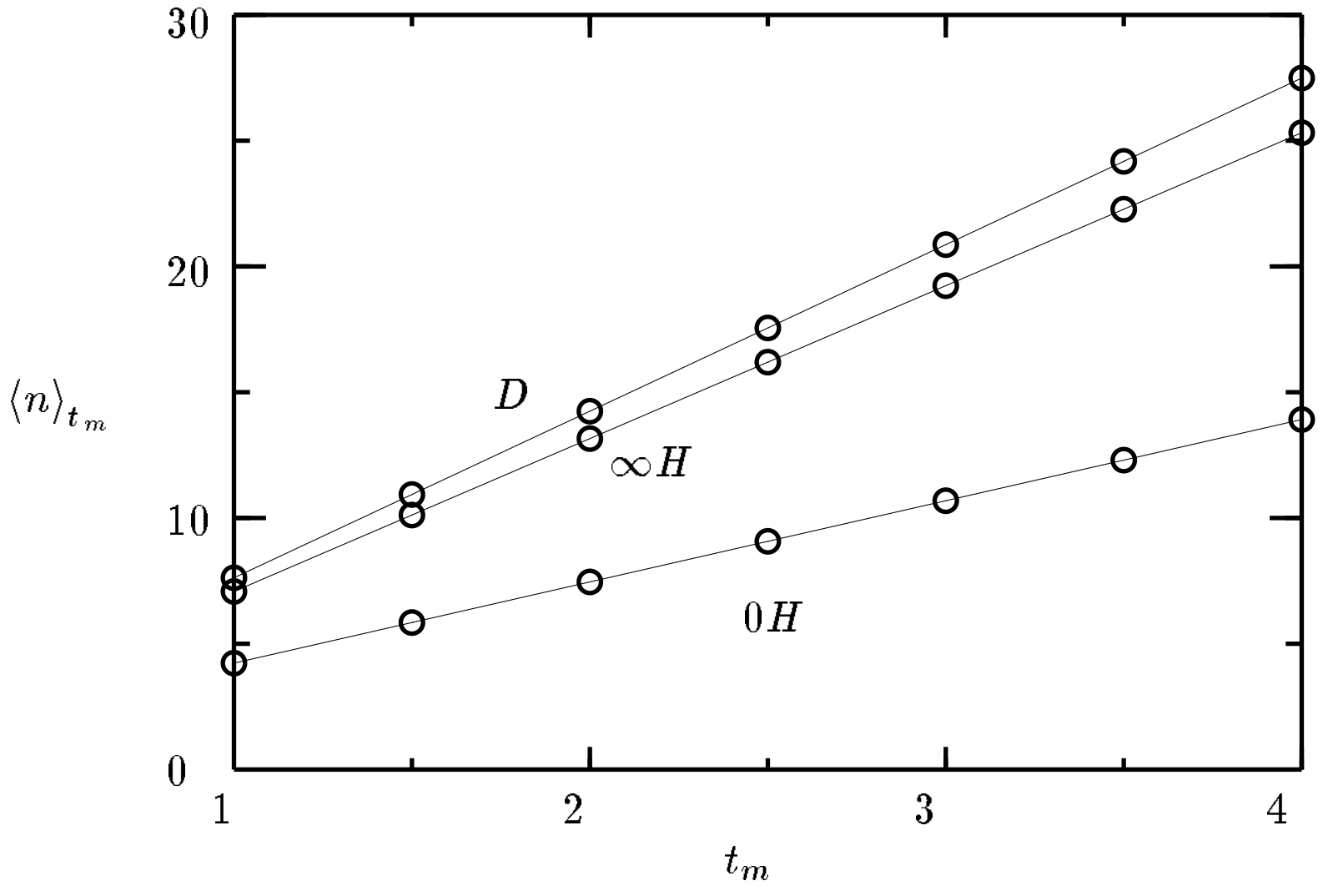

Fig. 23

*The average number of collisions by trajectory, $\langle n\rangle_{t_m}$, in a time interval $[0, t_m]$ for the $\infty H$ (3d), 0H and D cases. The dashed lines are the best linear fits of the data points.*

In **Fig.24** we show the logarithm of the distribution standard deviation (as computed from the data in the graphs in Fig. 19), $\ln\sigma_{t_m}$, for all three cases and different $t_m$. The fit $a_1 \ln(t_m) + a_2$ gives: $(a_1, a_2) = (0.4(\pm 0.1), -0.04(\pm 0.06))$ for the D case, (0.51 ($\pm$0.04),0.35 ($\pm$0.02)) for the $\infty$H case and (0.51 ($\pm$0.04), 0.35 ($\pm$0.02)) for the 0H case.

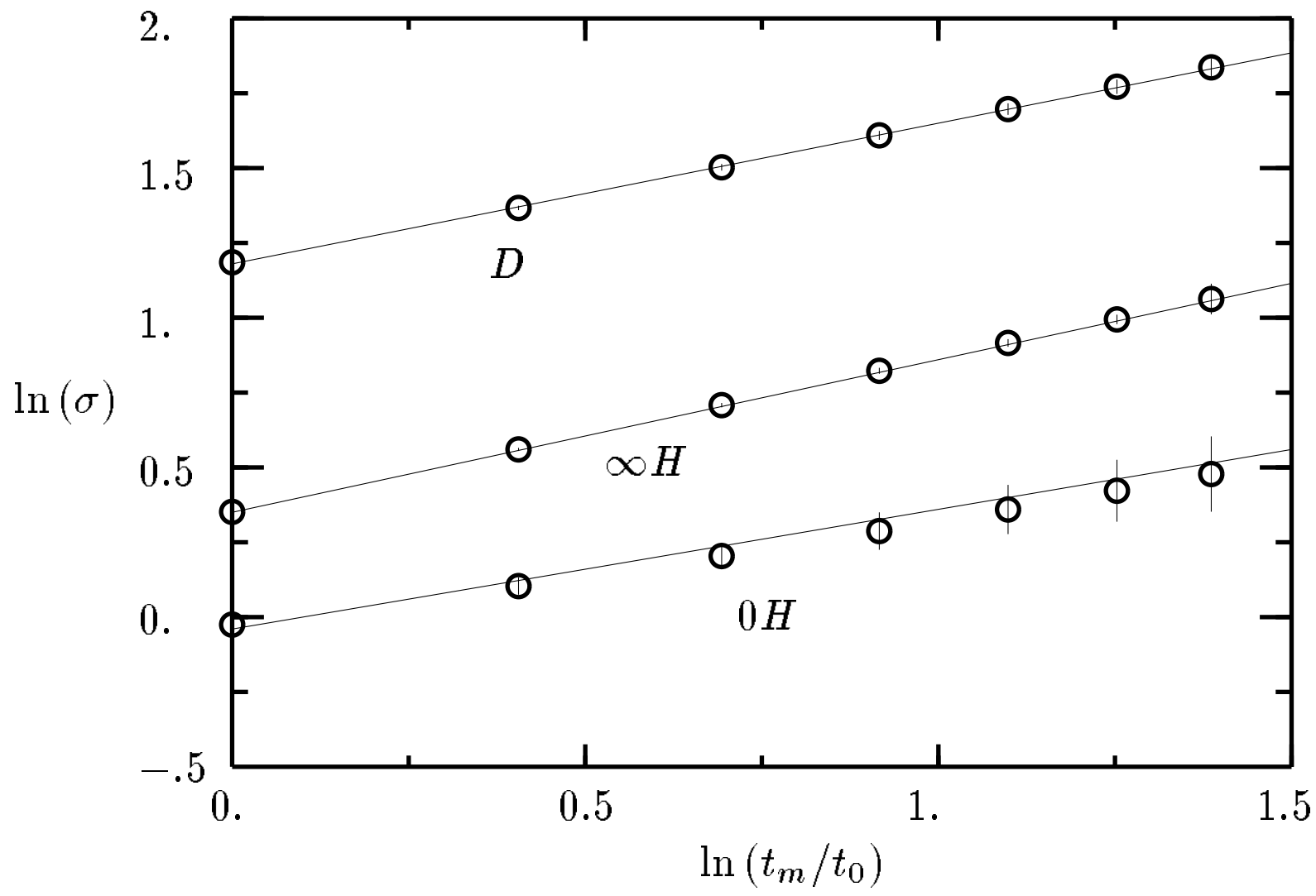

Fig. 24

*The logarithm of the standard deviation of the probability distribution $d(n; t_m)$ versus $\ln(t_m)$. The dashed lines are the best linear fit of the data points.*

From the above scaling information, it is natural to expect that the distribution of the random variable: $z = (n - \langle n\rangle_{t_m})/\sigma_{t_m}$ defined by $d_\alpha(z) = \sigma_{t_m} d(n(z); t_m)$ will be independent of $t_m$.

This is shown in **Fig.25**, **Fig.26** and **Fig.27** where we plotted all data for different $t_m$ in the $\infty$H (3$d$), 0H and $D$ cases respectively.

In the $\infty$H (3$d$) case, the scaled distribution, $d_\alpha(z)$, is not symmetric around the maximum. The distribution right wing (i.e. from the maximum to $n \to \infty$) seems to have a pure exponential behaviour. The left wing has a very fast decay. But it seems that the distribution depends on $t_m$ indicating that we are far from the asymptotic regime (which might still be a gaussian). For the 0H and $D$ cases the scaled data distribution can be fitted by a gaussian distribution: as expected from the rigorous theory, see [BS], [BSC].

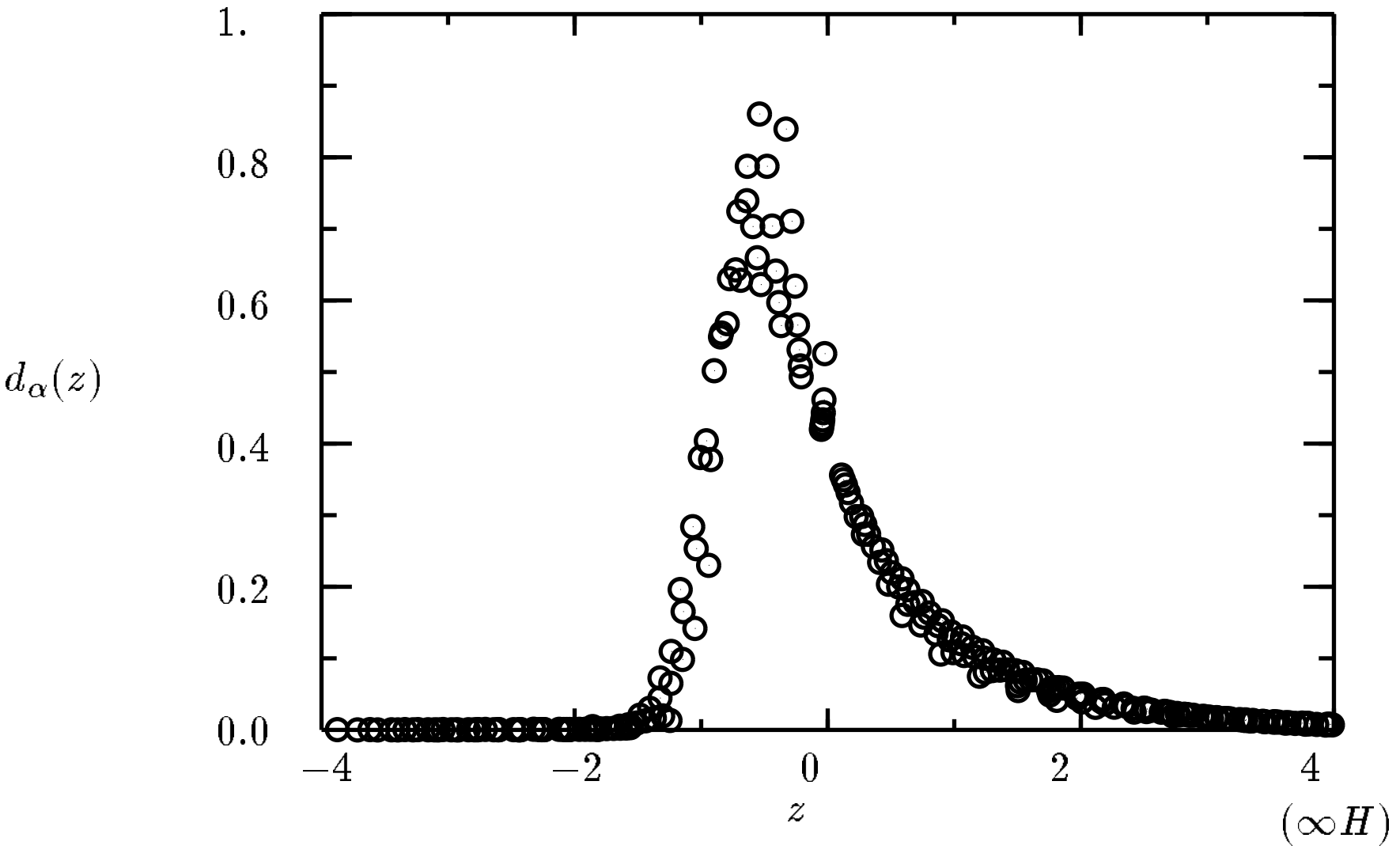

Fig. 25

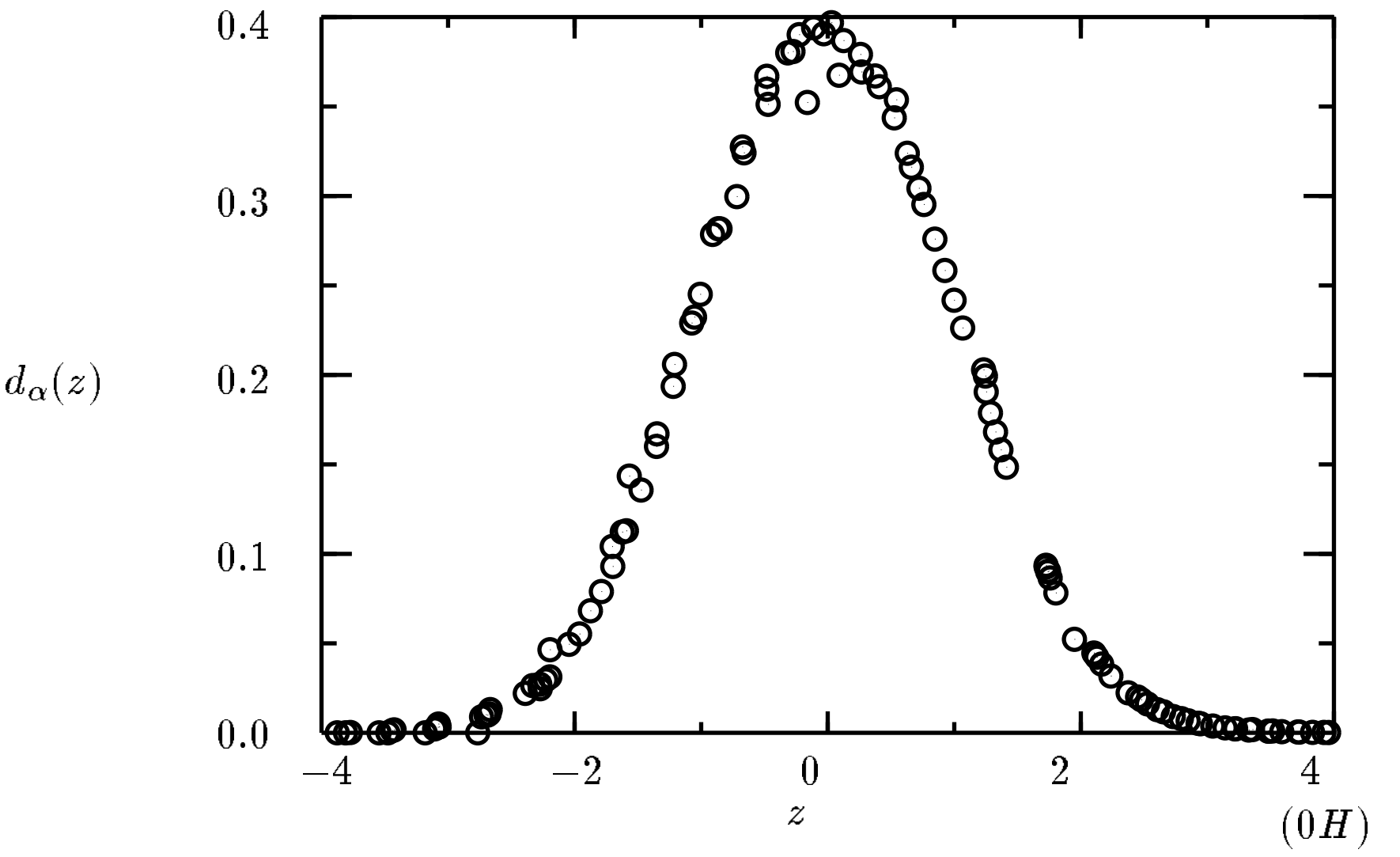

Fig. 26

*Scaled probability distribution:* $d_\alpha(z) = \sigma_{t_m} d(\langle n \rangle_{t_m} + \sigma_{t_m} z; t_m)$ *with different* $t_m$ *for the* $\infty H$ *(Fig.25) and* $0H$ *(Fig.26) cases.* $\langle z \rangle$ *is the average values of the scaled variable and* $\sigma$ *its standard deviation.*

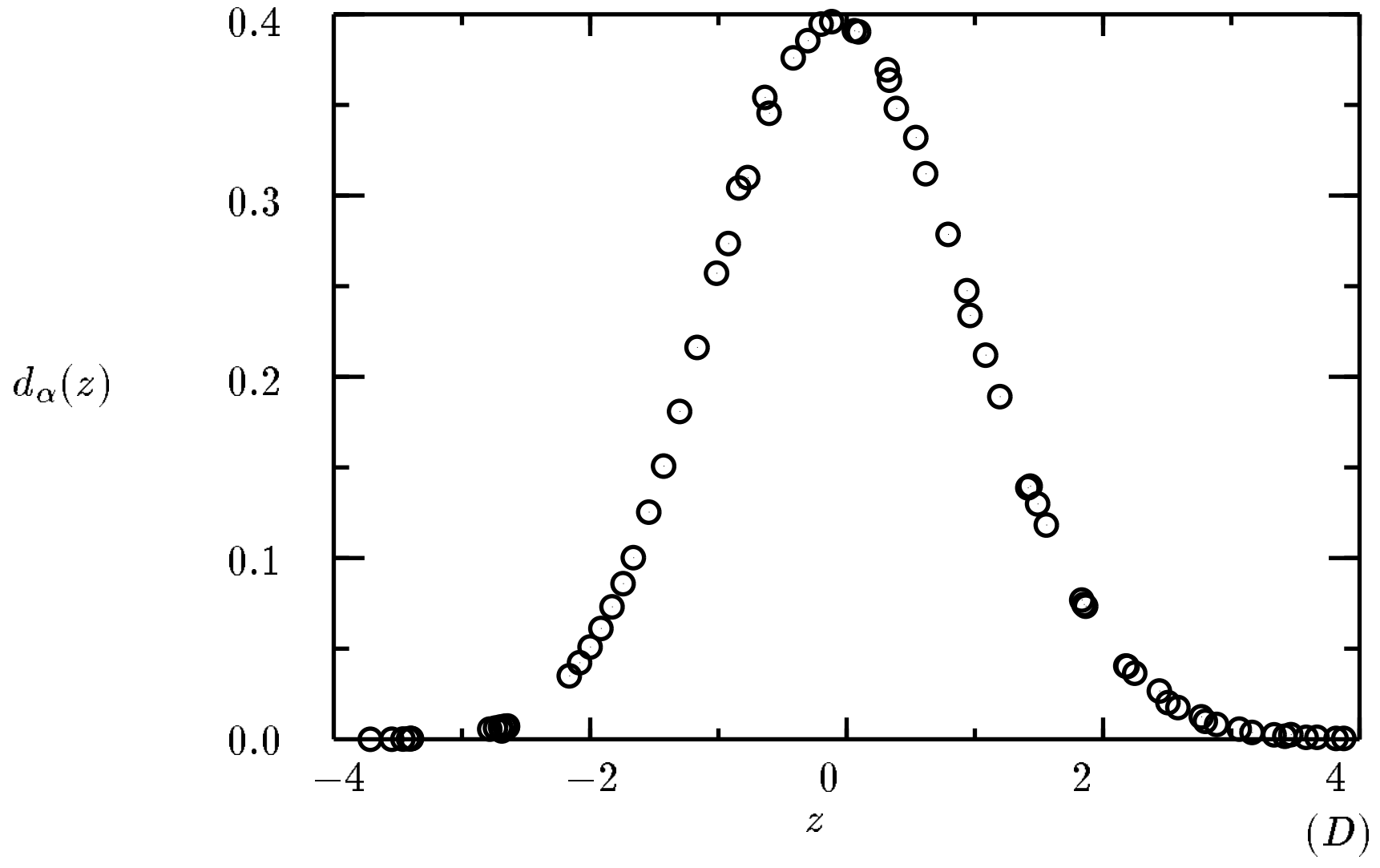

Fig. 27

*Scaled probability distribution:* $d_\alpha(z) = \sigma_{t_m} d(\langle n\rangle_{t_m} + \sigma_{t_m} z; t_m)$ *with different* $t_m$ *for the D case.* $\langle z\rangle$ *is the average values of the scaled variable and* $\sigma$ *its standard deviation.*

§*Appendix. Fits and Errors. The random number generator.*

From our computer experiments we get data sets corresponding to averages over many dynamical trajectories, say $\vec{y}(\vec{x}) = \{y(x_i)\}_{i=1,N}$, where $\vec{x} = \{x_i\}_{i=1,N}$ is in our case an independent variable, say for instance a set of $N$ collision numbers or time instants. To the latter experimental data set of points, we want to *fit* a given *guessed* function, say $f(x;\vec{\alpha})$, where $\vec{\alpha} = \{\alpha_n\}_{n=1,p}$ is a set of arbitrary parameters. Here by fit we mean to find a set of parameters $\vec{\alpha}^*$ which *optimizes* some reasonable functional relation between the experimental data and the fitting function.
In our case we use the *least squares funtional, i.e.* :

$$V(\vec{y}(x),\vec{\alpha}) = \sum_{i=1}^{N} \left[y_i - f(x_i;\vec{\alpha})\right]^2 \qquad (A1.1)$$

The set of parameters $\vec{\alpha}^*(\vec{y})$ is here obtained by asking that they should be the minima of the $V$ function: $\partial_{\vec{\alpha}^*} V(\vec{y},\vec{\alpha}^*) = 0$ . We also define the *goodness* of our fit, $G$, by the average $y$-distance of our data to the function $f(x;\vec{\alpha}^*)$: $G(\vec{y}(\vec{x})) = V(\vec{y}(\vec{x}),\vec{\alpha}^*)/N$. This parameter is only meaningful when it is compared with the one from another fit. Given many fits, the one with smallest $G$ value will be called *best fit* (among the considered fits).
The data have, in general, non-negligible errors, say $\vec{\varepsilon} = \{\varepsilon_i\}_{i=1,N}$, due to the finite number of samples used in the averaging and to the rounding error propagation because of the system chaoticity (see comments in §2). Such errors induce errors on the parameter values. Therefore, a measure of the error amplitude in $\vec{\alpha}^*(\vec{y})$ is given by:

$$\Delta_{\vec{\varepsilon}}\vec{\alpha}^*(\vec{y}) = \left[\vec{\alpha}^*(\vec{y}+\vec{\varepsilon}) - \vec{\alpha}^*(\vec{y}-\vec{\varepsilon})\right]/2 \qquad (A1.2)$$

In the particular case in which the magnitude of the data error is much smaller than the measured value, $|\varepsilon_i/y(x_i)| << 1$, we may expand the latter equation around $\vec{\varepsilon} = \vec{0}$:

$$\Delta_{\vec{\varepsilon}}\alpha_n(\vec{y}) = \sum_{i=1}^{N} c_i^{(n)}(\vec{y})\varepsilon_i \qquad (A1.3)$$

The coefficients $c_i^{(n)}$ are found by expanding $V(\vec{y}(\vec{x})+\vec{\varepsilon},\vec{\alpha})$ around $\vec{\alpha}^*(\vec{y})$ and $\vec{\varepsilon} = \vec{0}$ and they are given by:

$$c_i^{(n)} = -\sum_{m=1}^{p} D^{-1}_{mn}\,\partial^2_{y(x_i)\alpha^*_m} V(\vec{y},\vec{\alpha}^*) \qquad (A1.4)$$

where $D_{mn} = \partial^2_{\alpha^*_m \alpha^*_n} V(\vec{y}, \vec{\alpha}^*)$.

In particular for the linear fit, $f(x, \vec{\alpha}) = \alpha_1 + \alpha_2 x$, the coefficients $c_i^{(1),(2)}$ are given by:

$$c_i^{(1)} = \frac{2}{N \Delta x}(\overline{x^2} - \overline{x} x_i), \qquad c_i^{(2)} = \frac{2}{N \Delta x}(x_i - \overline{x}) \tag{A1.5}$$

where $\Delta x = \overline{(x - \overline{x})^2}$ and $\overline{x^n} = N^{-1} \sum_{i=1}^N x_i^n$.

The errors are random variables and we have to average them over their distribution. Since our data we have come from dynamical trajectories, the errors may be correlated and we cannot assume that they are independent gaussian distributed random variables. Therefore we *empirically* estimate an upper bound for their correlation values:

$$|\langle \varepsilon_i \varepsilon_j \rangle| \leq \sqrt{\langle \varepsilon_i^2 \rangle \langle \varepsilon_j^2 \rangle} e^{-|x_i - x_j| \lambda} \tag{A1.6}$$

where $\lambda$ is the corresponding Lyapunov exponent. The parameter errors in our analysis are defined by the equations:

$$\Delta \alpha_n(\vec{y})^2 = \sum_{i=1}^N \sum_{j=1}^N c_i^{(n)} c_j^{(n)} \delta_i \delta_j e^{-\lambda |x_i - x_j|} \geq \langle \Delta_{\vec{\varepsilon}} \alpha_n^2 \rangle \tag{A1.7}$$

where $\delta_i^2 = \langle \varepsilon_i^2 \rangle$ and their use and meaning is described entirely by the above comments.

Finally a comment on our random number generator: we have used the so called R250 in which a sequence of pseudo-random numbers, $\{X(n)\}$, is generated by the linear recursion relation: $X(n) = X(n-103).xor.X(n-250)$ where *.xor.* is the logical operation *exclusive or* along the bits of both numbers, *i.e.* $1.xor.1 = 0$, $0.xor.0 = 0$ and $1.xor.0 = 1$. The first 250 random numbers are generated with a random number *modulo generator*: $X(n) = 16807 * X(n-1) \bmod (2^{31} - 1)$. The recurrence period for the R250 random number is expected to be $(2^{103} - 1) * (2^{250} - 1)$. We do not have reasons to think that the random numbers could be "bad"; our results should be reproducible, although their interpretation might be different from ours.